\documentclass[aps,prd,preprint,groupedaddress,amssymb,amsmath,nofootinbib]{revtex4}

\usepackage{color}
\usepackage{graphicx}
\usepackage{epstopdf}

\usepackage{slashed} 

\begin{document}

\preprint{}

\title{Explanation of the Muon g-2 Anomaly with Vectorlike Leptons \\ and its Implications for  Higgs Decays}

\author{Radovan Derm\' \i\v sek}
\email[]{dermisek@indiana.edu}
\affiliation{Physics Department, Indiana University, Bloomington, IN 47405, USA}

\author{Aditi Raval}
\email[]{adiraval@indiana.edu}
\affiliation{Physics Department, Indiana University, Bloomington, IN 47405, USA}


\date{May 30, 2013}

\begin{abstract}

The deviation of the measured value of the muon anomalous magnetic moment from the standard model prediction can be completely explained by mixing of the muon with extra vectorlike leptons, L and E,  near the electroweak scale. This mixing simultaneously contributes to the muon mass. We show that the correlation between contributions to the muon mass and muon g-2 is controlled by the mass of the neutrino originating from the  doublet L. Positive correlation, simultaneously explaining both measured values, requires this  mass  below 200 GeV. 
The decay rate of the Higgs boson to muon pairs  is modified and, in the region of the parameter space that can explain the muon anomalous magnetic moment within one standard deviation, it ranges from 0.5 to 24 times the standard model prediction.
In the same scenario, 
 $h \to \gamma \gamma$
can be enhanced  or lowered by $\sim$50\% from the standard model prediction. The explanation of the muon g-2 anomaly and predictions for $h \to \gamma \gamma$ are not correlated since these are controlled by independent parameters. 
This scenario can be embedded  in a model with three complete vectorlike families featuring gauge coupling unification, sufficiently stable proton, and the Higgs quartic coupling remaining positive all the way to the grand unification scale.

\end{abstract}


\maketitle






 \section{Introduction}

 The measured value of the muon anomalous magnetic moment represents one of the largest discrepancies from predictions of the standard model (SM). There has been a variety of new physics models attempting to explain this deviation~\cite{Jegerlehner:2009ry}. Most of the effort has been within the frameworks related to the explanation of the hierarchy between the electroweak (EW) scale and the grand unification (GUT) scale or the Planck scale. 
  
 However, we continue to see no signs of new physics related to solving the hierarchy problem at the LHC, and many well motivated possible explanations of the muon g-2 anomaly are now excluded. This motivates us to step back and see whether there are other simple ways to explain the anomaly, that are testable at the LHC, but not necessarily related to the naturalness problem of the electroweak symmetry breaking.
 
 In this paper we show that the deviation of the measured value of the muon anomalous magnetic moment from the standard model prediction can be completely explained by mixing of the muon with extra vectorlike leptons, $L$  and $E$,  near the electroweak scale. This mixing simultaneously contributes to the muon mass.  We find that the correlation between contributions to the muon mass and muon g-2 is controlled by the mass of the neutrino originating from the  doublet L,  that is given by the vectorlike mass parameter $M_L$. Positive correlation, simultaneously explaining both measured values, requires this parameter to be small,  $M_L \lesssim 200$ GeV. We further discuss implication of this scenario for Higgs decays, namely $h \to \mu^+ \mu^-$ and $h \to \gamma \gamma$, and provide a UV embedding of this scenario  with many attractive features.

The possibility of explaining the muon g-2 anomaly by mixing of the muon with extra heavy leptons  was previously noticed  in  Refs.~\cite{Czarnecki:2001pv, Kannike:2011ng}. The mass enhancement  originating from the loop involving a heavy lepton is compensated by a small flavor violating couplings (which originate from the mixing and thus they are suppressed by the mass of the heavy lepton). Therefore, the new physics contribution to the muon g-2, with new fermions at or below the TeV scale, can be of the same order as the contributions of the $W$ and $Z$ bosons in the standard model. Similar effect can be obtained with a $Z'$ with flavor violating couplings of the muon to  a heavier lepton, see for example Ref.~\cite{Murakami:2001cs}. 

Indeed, in several scenarios with new leptons near the EW scale explored recently in Ref.~\cite{Kannike:2011ng}, it was found that the size of the muon g-2 anomaly is naturally of the same order as the contribution generated by heavy leptons,  when their contribution to the muon mass is comparable to the physical muon mass.  However, it was found that a positive contribution to the muon mass results in a negative contribution to the muon g-2 and vice versa. This was explored in the asymptotic limit of taking masses of extra leptons large while keeping the mixing with the muon constant (by increasing Yukawa couplings). 
We will show that this anticorrelation only happens in the asymptotic limit due to the dominance of the Higgs contribution.  For smaller masses, it is the $W$ contribution, controlled by $M_L$,  that dominates. This reverses the sign of the correlation and a simultaneous explanation of the muon mass and muon g-2 from the mixing can be achieved, and it is fairly generic for small $M_L$.\footnote{This possibility was missed in the original version of Ref.~\cite{Kannike:2011ng}, arXiv:1111.2551v1 [hep-ph], as a result of a mistake in the calculation of the $W$ contribution. It was pointed out by one of the authors of this paper, A.R.. In the corrected version of Ref.~\cite{Kannike:2011ng} some points  with a positive correlation between contributions to the muon g-2 and muon mass appeared, but the focus of the paper remained on the asymptotic case.} In addition, an arbitrary correlation can be achieved in between the small $M_L$ case, dominated by the $W$ loop, and the asymptotic case, dominated by the Higgs loop.

 Mixing of the muon with heavy leptons generically leads to a modification of the Higgs coupling to the muon~\cite{Giudice:2008uua}~\cite{Kannike:2011ng}.  
 Thus, the decay rate of the Higgs boson to muon pairs  is modified, and in the region of the parameter space that can explain the muon g-2 within $1\sigma$, it 
  ranges from $\sim$0.5 to $\sim$24 times the standard model prediction. A part of the  parameter space is already excluded by the ATLAS search for $h \to \mu^+ \mu^-$, that with 
 20.7 fb$^{-1}$ collected at 8 TeV sets the limit 9.8  times the SM prediction~\cite{Atlas_h_to_mumu}.
  
  The scenario also allows for a sizable modification of  $h \to \gamma \gamma$, since extra charged leptons can appear in loops mediating this process. This was recently extensively discussed in  Refs.~\cite{Joglekar:2012vc, ArkaniHamed:2012kq, Almeida:2012bq, Kearney:2012zi}, motivated by the observed rate for $h \to \gamma \gamma$ at both the ATLAS and CMS  experiments being significantly above the SM prediction at some point. However, with more data collected, the current ATLAS result is $1.65 \pm 0.35$ times the SM prediction~\cite{ATLAS_h_gamma_gamma}, while the CMS experiment finds $0.78\pm0.27$~\cite{CMS_h_gamma_gamma}.
In the region of the parameter space that can explain the muon g-2 within $1\sigma$, limiting the size of Yukawa couplings to 0.5, motivated by a simple UV embedding, 
the branching ratio for $h \to \gamma \gamma$
can be enhanced by   $\sim$15\% or lowered by $\sim$25\%. Allowing Yukawa couplings of order 1 the $h \to \gamma \gamma$ rate can be modified by $\sim$50\%.
The explanation of the muon g-2 anomaly and predictions for $h \to \gamma \gamma$ are however not correlated, since these are controlled by independent parameters.

Models with flavor violating couplings are typically highly constrained by limits on a variety of flavor changing precesses. However, these constraints involve products of flavor violating couplings of two different light leptons, while for the explanation of the muon g-2 anomaly only the couplings of the muon to heavy leptons are necessary. We can therefore take the existing limits on flavor violating processes as constraints on other couplings in the model that are not necessary for the explanation of the muon g-2 anomaly.

 While the extra vectorlike leptons, $L$  and $E$, that mix with the muon are sufficient for the explanation of the muon g-2 anomaly, this does not have to be the full story. The model can be combined with other scenarios involving vectorlike fermions. For example,  it is possible to embed it 
  into recently discussed scenario with extra 3 or more complete  vectorlike families~\cite{Dermisek:2012as, Dermisek:2012ke}
  featuring gauge coupling unification, sufficiently stable proton, and the Higgs quartic coupling remaining positive all the way to the GUT scale. 
In this scenario, predicted values of gauge couplings at the electroweak scale are highly insensitive to GUT scale parameters and masses of vectorlike fermions. They can be understood from  IR fixed point predictions and threshold effects from integrating out vectorlike families. Furthermore, a model with extra $Z'$ and vector-like quarks was recently discussed as a possible explanation of the anomalies in  $Z$-pole observables: the forward-backward asymmetry of the b-quark, and the
lepton asymmetry obtained from the measurement of left-right asymmetry for hadronic final states~\cite{Dermisek:2011xu, Dermisek:2012qx}. These two scenarios could also be combined, and a simultaneous explanation of anomalies in $Z$-pole observables and the muon g-2 could be obtained. 
  
 This paper is organized as follows. In Sec.~\ref{sec:model}, we define the model and find expressions for couplings of $Z$, $W$ and $h$ to the SM and extra leptons.
 In Sec.~\ref{sec:g-2}, we calculate contributions of extra leptons to the muon anomalous magnetic moment, qualitatively discuss  expected results, provide results from numerical scans over the parameters space, and discuss the predictions from the regions that explain the muon g-2 for higgs decays, namely $h \to \mu^+ \mu^-$ and $h \to \gamma \gamma$. We also discuss constraints from precision EW observables, current constraints from the LHC, and encourage further searches for extra leptons in a variety of final states at the LHC.
In Sec.~\ref{sec:SM+3VFs}, we discuss a possible UV embedding of this model in the extension of the SM with three complete vector like families.  We provide some concluding remarks in Sec.~\ref{sec:conclusions}.

 \section{Model}
 \label{sec:model}
 
 Quantum numbers of SM particles and extra vectorlike family (VF) are summarized in Table~\ref{tab:charges}. The notation is straightforward, we use lower case letters for standard model particles and upper case letters for particles from extra VF, {\it e.g.} $E_R$ has the same quantum numbers as $e_R$ and its vector like partner is $E_L$.
For the discussion of the muon g-2 only $L_{L,R}$ and $E_{L,R}$ are relevant. Extra quarks obviously do not contribute and we will not assume that the standard model singlets $N_{L,R}$  are near the EW scale.

\begin{table}[h]
    \caption{Quantum numbers of  standard model and extra vectorlike particles. The electric charge is given by $Q = T_3 +Y$, where $T_3$ is the weak isospin, which is +1/2 for for the first component of a doublet and -1/2 for the second component. }
    
\begin{tabular}{l c c c c c c c | c c c c c c }
 \hline
  \hline
   & $q_L$ & $u_R$ & $d_R$ & $l_L$ & $\nu_R$ & $e_R$ & $H$ & $Q_{L,R}$ & $U_{L,R}$ & $D_{L,R}$ & $L_{L,R}$ & $N_{L,R}$ & $E_{L,R}$ \\
 \hline
SU(3)$_{\text C}$ & {\bf 3} & {\bf 3} & {\bf 3}  &  {\bf 1} &  {\bf 1} & {\bf 1} & {\bf 1} & {\bf 3} & {\bf 3} & {\bf 3} &  {\bf 1} &  {\bf 1} & {\bf 1} \\
SU(2)$_{\text L}$ & {\bf 2} & {\bf 1} & {\bf 1} & {\bf 2} & {\bf 1}  & {\bf 1} & {\bf 2} & {\bf 2} & {\bf 1} &{\bf 1} & {\bf 2} &  {\bf 1} & {\bf 1} \\
U(1)$_{\text Y}$   & $\frac{1}{6}$ & $\frac{2}{3}$ & -$\frac{1}{3}$ & -$\frac{1}{2}$ & 0 & -1 &$\frac{1}{2}$& $\frac{1}{6}$ & $\frac{2}{3}$ & -$\frac{1}{3}$ & -$\frac{1}{2}$ & 0 &  -1 \\
        \hline
  \hline
      \end{tabular}
   \label{tab:charges}
\end{table}

The most general renormalizable lagrangian for charged leptons is given by:
\begin{eqnarray}
{\cal L} &\supset& -  \bar l_{Li} y_{ij} e_{Rj} H - \bar l_{Li} \lambda^E_i E_{R} H   - \bar L_{L} \lambda^L_j e_{Rj} H -  \lambda \bar L_{L}  E_{R} H - \bar \lambda H^\dagger \bar E_{L}  L_{R}  \nonumber \\ 
&& - M_L \bar L_L L_R - M_E \bar E_L E_R + {\it h.c.},
\end{eqnarray}
where the terms in the first line represent the usual standard model Yukawa couplings, (the sum over flavor indices is assumed), Yukawa couplings between SM leptons and leptons from VF, and between leptons from VF. Terms in the second line are mass terms for vectorlike pairs of leptons. 
We label the components of doublets as follows:
\begin{equation}
l_{Li} = \left(  \begin{array}{c}   
\nu_i \\
e_{Li}
\end{array}
 \right), \quad \quad \quad 
 L_{L,R} = \left(  \begin{array}{c}   
L^0_{L,R} \\
L^-_{L,R}
\end{array}
 \right), \quad \quad \quad 
H = \left(  \begin{array}{c}   
0 \\
v + \frac{h}{\sqrt{2}}
\end{array}
 \right),
\end{equation}
where $v = 174$ GeV is the vacuum expectation value of the Higgs field.  

After spontaneous symmetry breaking the $5 \times 5$ mass matrix for charged leptons is given by:
\begin{eqnarray}
( \bar e_{Li}, \bar L^-_L, \bar E_L ) \; 
M_e \;  \begin{pmatrix}
 e_{Rj} \\
 L^-_R\\
 E_R
\end{pmatrix}
= 
( \bar e_{Li}, \bar L^-_L, \bar E_L ) 
\begin{pmatrix}
 y_{ij} v & 0 &  \lambda^E_{i} v\\
  \lambda^L_{j} v & M_L &  \lambda v\\
 0 & \bar \lambda v & M_E 
\end{pmatrix}
\begin{pmatrix}
 e_{Rj} \\
 L^-_R\\
 E_R
\end{pmatrix},
\end{eqnarray}
and it is convenient to define 5-component vectors: $e_{La} \equiv (e_{Li}, L_L^-, E_L)^T$ (and similarly for $e_{Ra}$ with $L\rightarrow R$), which combine the left (right) handed fields of the SM with those from the extra vectorlike pairs. We use indices from the beginning of the alphabet  for combined vectors and indices starting with $i$ for only the standard  model leptons. 
This mass matrix  can be diagonalized by a bi-unitary transformation, $U^\dagger_L M_e U_R$, which defines the mass eigenstate basis. 
We label the mass eigenstates by $e_a$ and for the lightest three eigenstates we will also use their names: $e$, $\mu$ and $\tau$.

Before diagonalizing the full mass matrix,  it is instructive to  
change the basis by a unitary transformation, $e_{Li} \to (V_L e_L)_i$, $e_{Rj} \to (V_R e_R)_j$,   which diagonalizes the standard model Yukawa couplings $y_{ij}$. The  mass matrix becomes
\begin{eqnarray}
\begin{pmatrix}
 (V^\dagger_{L} y V_{R} )_{ij} v & 0 &  (V^\dagger_{L})_{ik} \lambda^E_{k} v\\
  \lambda^L_{l} (V_{R})_{lj} v & M_L &  \lambda v\\
 0 & \bar \lambda v & M_E \\
\end{pmatrix}.
\end{eqnarray}
Since we are interested in modifying couplings of the muon, we assume that 
only $(V^\dagger_{L})_{2k} \lambda^E_{k}$ and  $\lambda^L_{l} (V_{R})_{l2}$ are non-zero. 
 This corresponds to the situation when $\lambda^E_{k} \propto (V_{L})_{k2} $ and $\lambda^L_l \propto (V_{R}^\dagger)_{2l}$, or in the basis where standard model Yukawa couplings are diagonal, it corresponds to $ \lambda^{L,E}_1 = \lambda^{L,E}_3 = 0$ and $\lambda^{L,E}_2 \equiv \lambda^{L,E}$ is non-zero.   This is the minimal scenario that does not modify standard model couplings of the electron and tau. 

In this minimal scenario, masses of the electron and tau fully originate from their Yukawa couplings to the Higgs boson since they do not mix with heavy leptons. Therefore we can look at the $3 \times 3$ mass matrix for the muon and the extra heavy leptons separately:
\begin{eqnarray}
U^\dagger_L
\begin{pmatrix}
 y_\mu v & 0 &   \lambda^E v\\
  \lambda^L v & M_L &  \lambda v\\
 0 & \bar \lambda v & M_E \\
\end{pmatrix}
U_R 
 =  
 \begin{pmatrix}
m_\mu  & 0 &   0\\
 0 & m_{e_4} &  0\\
 0 & 0 & m_{e_5} \\
\end{pmatrix},
\label{eq:mass}
\end{eqnarray}
 where we use the same names for diagonalization matrices $U_{L,R}$ as for the matrices  that diagonalize the general $5 \times 5$  matrix.
 We label their components  by 2, 4 and 5 so that results are applicable to the general scenario. Similarly we label the heavy mass eigenstates by $e_4$ and $e_5$.

 In the limit   \begin{equation}
\lambda_E v, \lambda_L v, \bar \lambda v, \lambda v  \ll M_E, M_L,
\label{eq:integrateout}
 \end{equation}
approximate analytic formulas for diagonalization matrices can be obtained which are useful for deriving approximate formulas for  couplings of $Z, W$ and $h$. In this limit, the two heavy charged leptons have masses close to $ M_L$ and $M_E$. In the basis $(m_\mu, m_{e_4} \simeq M_L, m_{e_5} \simeq M_E)$ the diagonalization matrices are given by:
\begin{equation}
U_{L} =  \renewcommand\arraystretch{1.5}  \left(\begin{array}{ccc}
1 - v^2 \frac{\lambda_E^2}{2M_E^2}  						& 	-v^2 \left( \frac{\lambda_E}{M_L}\frac{ \bar{\lambda} M_E + \lambda M_L}{M_E^2 - M_L^2} - \frac{y_\mu \lambda_L}{M_L^2}\right)				& 	v\frac{\lambda_E}{M_E} 	   \\
 v^2 \frac{ \bar{\lambda} \lambda_E M_L  - y_\mu \lambda_L M_E  }{M_L^2 M_E} 		& 	1 - v^2 \frac{(\lambda M_E +  \bar{\lambda} M_L)^2}{2(M_E^2 - M_L^2)^2}   &  v \frac{ \bar{\lambda} M_L + \lambda M_E}{M_E^2 - M_L^2} \\
- v\frac{\lambda_E}{M_E}  &     -v \frac{ \bar{\lambda} M_L + \lambda M_E}{M_E^2 - M_L^2}					&   1  -  v^2 \frac{\lambda_E^2}{2 M_E^2}  - v^2\frac{(\lambda M_E +  \bar{\lambda} M_L)^2} {2(M_E^2 - M_L^2)^2}    \end{array}   \right)  
\label{eq:UL}
 \end{equation}
and
\begin{equation} 
U_R =  \renewcommand\arraystretch{1.9} \left(\begin{array}{ccc}
1 - v^2 \frac{\lambda_L^2}{2M_L^2}			   & 	v \frac{\lambda_L}{M_L} 										& 	v^2 \left(\frac{\lambda_L}{M_E} \frac{ \bar{\lambda} M_L + \lambda M_E}{ M_E^2 - M_L^2} + \frac{y_\mu \lambda_E}{M_E^2}  \right)   \\
  -v\frac{\lambda_L}{M_L}   & 	1 - v^2\frac{ \lambda_L^2}{2 M_L^2} - v^2\frac{( \bar{\lambda} M_E + \lambda M_L)^2}{2 (M_E^2 - M_L^2)^2}  	&   v\frac{ \bar{\lambda} M_E + \lambda M_L} {M_E^2 - M_L^2}   \\
    v^2\frac{ \lambda_L  \bar{\lambda} M_E  -  y_\mu \lambda_E M_L}{M_LM_E^2} 	   					   &   -v\frac{ \bar{\lambda} M_E + \lambda M_L}{M_E^2 - M_L^2} 						&  1- v^2\frac{( \bar{\lambda} M_E + \lambda M_L)^2}{2 (M_E^2 - M_L^2)^2}  					
\end{array}   \right)  .
\label{eq:UR}
\end{equation}

\subsection{Couplings of the $Z$ boson and photon}

Couplings of the electron and $\tau$ to the $Z$ boson  are not modified from their SM values.
Couplings of  other charged leptons  to the Z boson are modified because the extra $E_L$  is an SU(2) singlet but it mixes with SU(2) doublets, and $L_R^-$ originates from an SU(2) doublet but it mixes with SU(2) singlets. The couplings follow from  the kinetic terms:
\begin{eqnarray}
{\cal L}_{kin} \supset   \bar e_{La} i \slashed D_a e_{La} +  \bar e_{Ra} i \slashed D_a e_{Ra}  \; = \; \bar {\hat e}_{L a} (U^\dagger_L)_{a c} i \slashed D_c (U_L)_{ cb} \hat e_{L b} +  \bar {\hat e}_{R a} (U^\dagger_R)_{a c} i \slashed D_c (U_R)_{c b } \hat e_{R b},
\label{eq:kinE}
\end{eqnarray}
where the vectors of mass eigenstates are $\hat e_{La} \equiv (\hat \mu_L,\hat e_{L4}, \hat e_{L5})^T$ and similarly for $\hat e_{R}$.
The covariant derivative is given by:
\begin{eqnarray}
D_{\mu a} = \partial _\mu 
- i \frac{g}{\cos \theta_W} (T^3_a - \sin ^2 \theta_W Q_a) Z_\mu - i e Q_a A_\mu,
\end{eqnarray}
where the weak isospin $T^3$ and the electric charge $Q$  for a given field can be obtained from Table~\ref{tab:charges}.
Defining couplings of the Z boson  to fermions $f_a$ and $f_b$ by the lagrangian of the form
\begin{equation}
{\cal L} \supset      \left( \bar f_{La} \gamma^\mu g^{Zf_a f_b}_L  f_{Lb} + \bar f_{Ra} \gamma^\mu g^{Zf_a f_b}_R  f_{Rb} \right) Z_\mu
\end{equation}
we find the couplings of left- and right-handed fields:
 \begin{eqnarray}
g^{Ze_a e_b}_{L} & = &\frac{g}{\cos \theta_W} \displaystyle\sum\limits_{c=2,4,5} (T^3_{Lc} - \sin ^2 \theta_W Q_c) (U^\dagger_L)_{ac} (U_L)_{cb}, 
\label{eq:gZL}\\
g^{Ze_a e_b}_{R} & = & \frac{g}{\cos \theta_W} \displaystyle\sum\limits_{c=2,4,5} (T^3_{Rc} - \sin ^2 \theta_W Q_c) (U^\dagger_R)_{ac} (U_R)_{cb},
\label{eq:gZR}
\end{eqnarray}
where $Q_c = -1$ is the same for all states;
$T^3_{Lc} = -1/2$ for $c = 2, 4$,  and $0$ for $c=5$; and $T^3_{Rc} = 0$ for $c = 2, 5$,  and $-1/2$ for $c=4$.

Since $Q_c = -1$ for all the fields, couplings of the photon are not modified  from their SM values by field rotations.
However, due to different weak isospins of fields that mix, the couplings of the  $Z$  boson given in Eqs.~(\ref{eq:gZL}) and (\ref{eq:gZR}) are modified. Corrections to the usual SM value of the left-handed charged lepton coupling,
 \begin{eqnarray}
(g^{Ze_a e_b}_L)_{SM} &=& \frac{g}{\cos \theta_W}  (-\frac{1}{2} + \sin ^2 \theta_W ) \delta^{ab}
\label{eq:gLSM}
\end{eqnarray}
can be written as:
 \begin{eqnarray}
\delta g^{Ze_a e_b}_L &=& \frac{g}{2 \cos \theta_W}  (U^\dagger_L)_{a5} (U_L)_{5b}.
\label{eq:delgL}
\end{eqnarray}
Similarly, corrections to the usual SM value of the right-handed charged lepton coupling,
 \begin{eqnarray}
(g^{Ze_a e_b}_R)_{SM} &=& \frac{g}{\cos \theta_W}   \sin ^2 \theta_W  \delta^{ab}
\label{eq:gRSM}
\end{eqnarray}
can be written as:
 \begin{eqnarray}
\delta g^{Ze_a e_b}_R &=& - \frac{g}{2 \cos \theta_W}  (U^\dagger_R)_{a4} (U_R)_{4b}.
\label{eq:delgR}
\end{eqnarray}

\subsection{Couplings of the $W$ boson}

The couplings of the electron and $\tau$ to the $W$ boson are not modified from their SM values.
Couplings of other charged leptons follow from  the kinetic terms:
\begin{eqnarray}
{\cal L}_{kin} &\supset&   \frac{g}{\sqrt{2}}  \left(  \bar \nu_{\mu} \gamma^\mu \mu_{L} + \bar L_{L}^0 \gamma^\mu L_{L}^- +   \bar L_{R}^0 \gamma^\mu L_{R}^-  \right)W^+_\mu + h.c.  \\
&=& \frac{g}{\sqrt{2}}  \left(  \bar \nu_{\mu} \gamma^\mu (U_L)_{2b} \hat e_{L b} + \bar L_{L}^0 \gamma^\mu (U_L)_{4b} \hat e_{L b} +   \bar L_{R}^0 \gamma^\mu (U_R)_{4b} \hat e_{R b}  \right)W^+_\mu + h.c.  .
\label{eq:kinW}
\end{eqnarray}
Defining couplings of the W boson  to neutrinos $\nu_a$ and charged leptons $\hat e_b$ by the lagrangian of the form
\begin{equation}
{\cal L} \supset      \left( \bar \nu_{La} \gamma^\mu g^{W\nu_a e_b}_L  \hat e_{Lb} + \bar \nu_{Ra} \gamma^\mu g^{W\nu_a e_b}_R  \hat e_{Rb} \right) W^+_\mu + h.c.  ,
\end{equation}
we find couplings of left- and right-handed fields:
 \begin{eqnarray}
 g^{W\nu_\mu e_b}_{L} =  \frac{g}{\sqrt{2}}   (U_L)_{2b},  \quad \quad \quad  g^{W\nu_4 e_b}_{L} =  \frac{g}{\sqrt{2}}   (U_L)_{4b}, 
\label{eq:gWL}\\
g^{W\nu_4 e_b}_{R} =  \frac{g}{\sqrt{2}}   (U_R)_{4b}.
\label{eq:gWR}
\end{eqnarray}

\subsection{Couplings of the Higgs boson}

In the minimal scenario that we are focusing on, couplings of the electron and tau  to the Higgs boson are given by the  SM relations, $\lambda_{e,\tau} = m_{e,\tau}/v$. This usual relation between the mass of a particle and its coupling to the Higgs boson does not apply to other charged leptons 
 as a consequence of explicit mass terms for vectorlike fermions, $M_E$ and $M_L$.
 The couplings of the Higgs boson for other charged leptons follow from  the Yukawa terms:
\begin{eqnarray}
{\cal L}_{Y} \supset  - \frac{1}{\sqrt{2}} \, \bar e_{La}  \, Y_{ab} \, e_{Rb} \, h  \; + \; h.c. \; = \; -\frac{1}{\sqrt{2}}\, \bar {\hat e}_{L a} (U^\dagger_L)_{a c}  \, Y_{cd} \,  (U_R)_{ db} \, \hat e_{R b}  \, h  \; + \; h.c.,
\end{eqnarray}
where 
\begin{eqnarray}
Y = 
\begin{pmatrix}
 y_\mu  & 0 &  \lambda^E_{i} \\
  \lambda^L_{j}  &0 &  \lambda \\
 0 & \bar \lambda  & 0
\end{pmatrix}.
\end{eqnarray}
Since the $Y$ matrix is not proportional to the mass matrix given in Eq.~(\ref{eq:mass}) the Higgs couplings are in general flavor violating. 
Defining couplings of the Higgs boson  to mass eigenstates fermions $f_a$ and $f_b$ by the lagrangian of the form
\begin{equation}
{\cal L} \supset    - \frac{1}{\sqrt{2}} \, \bar f_{La}  \, \lambda_{f_a f_b} \,  f_{Rb}  \, h + h.c.,
\end{equation}
we find:
 \begin{eqnarray}
\lambda_{e_a e_b} & = & \displaystyle\sum\limits_{c,d=2,4,5}  (U^\dagger_L)_{ac} Y_{cd} (U_R)_{db}.
\label{eq:lamab}
\end{eqnarray}
Noticing that $Y v = M_e - diag (0,M_L,M_E)$ the Higgs boson couplings to mass eigenstates  can be alternatively written as:
\begin{eqnarray}
\lambda_{e_a e_b} v = 
\begin{pmatrix}
 m_\mu  & 0 &  0 \\
  0  &m_{e_4}&  0 \\
 0 & 0  & m_{e_5}
\end{pmatrix} 
- U^\dagger_L
\begin{pmatrix}
 0  & 0 &  0 \\
  0  &M_L&  0 \\
 0 & 0  & M_E
\end{pmatrix} 
U_R,
\end{eqnarray}
where the first term comes from the usual SM relation between fermion masses and their couplings to the Higgs boson and the second term represents contributions from the $M_{L,E}$ terms.

 In the limit   (\ref{eq:integrateout}), the
approximate analytic formulas for  all the couplings of $Z, W$ and $h$ can be easily obtained from diagonalization matrices (\ref{eq:UL}) and (\ref{eq:UR}).

 \section{The muon anomalous magnetic moment}
 \label{sec:g-2}

 The discrepancy between the measured value of the muon anomalous magnetic moment~\cite{Bennett:2006fi} and the SM prediction,
  \begin{equation}
 \Delta a^{exp}_\mu	= a^{exp}_\mu - a^{SM}_\mu = 2.7 \pm 0.80 \times 10^{-9}	,
 \label{eq:del_a_mu}
 \end{equation}
 that we will use in our analysis, is the average of evaluations of this discrepancy reported by several groups:
$2.49 \pm 0.87 \times 10^{-9}$~\cite{Aoyama:2012wk}, $2.61 \pm 0.80 \times 10^{-9}$~\cite{Hagiwara:2011af}, and $2.87 \pm 0.80 \times 10^{-9}$~\cite{Davier:2010nc}. 
On average, the discrepancy is at the level of 3.4 standard deviations.

The contributions to the muon magnetic moment from extra fermions originate from the loop diagrams with the Higgs, Z and W bosons shown in Fig.~\ref{fig:g-2}.  
Our calculation of these contributions, presented below, agrees with the results in Refs.~\cite{Leveille:1977rc, Lynch:2001zs} and also in the revised version of Ref.~\cite{Kannike:2011ng}. For references to the original calculation of the $Z$, $W$, and $h$ contributions in the SM, see Ref.~\cite{Jegerlehner:2009ry}.
 
 \begin{figure}[t,h]
\includegraphics[width=6.in]{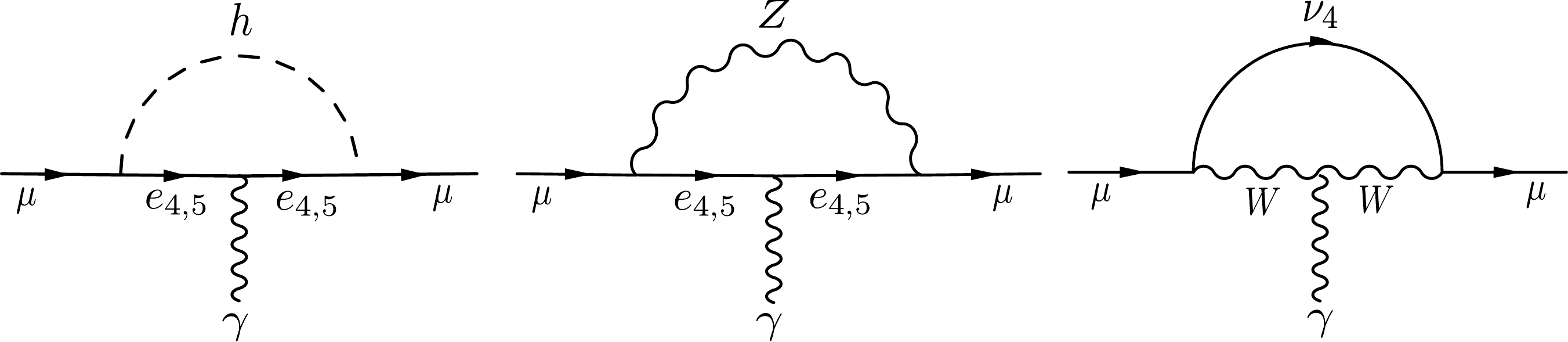} \caption{Feynman diagrams contributing to the muon magnetic moment that involve loops of extra fermions and the Higgs, Z and W bosons.}
\label{fig:g-2}
\end{figure}

The contribution from the Higgs diagram is given by
\begin{equation}
\delta a_\mu^h = -\frac{m_\mu}{32\pi^2 M_h^2} \sum_{b=4,5}  \left[(|\lambda_{\mu e_b}|^2 + |\lambda_{e_b \mu}|^2) \, m_\mu F_h(x_{hb}) +   {\rm Re}\, (\lambda_{\mu e_b}   \lambda_{e_b \mu} )  \, m_{e_b}  G_h(x_{hb})  \right] ,
\end{equation}
where $x_{hb} \equiv (m_{e_b}/M_h)^2$, the couplings are given in Eq.~(\ref{eq:lamab}) with index $\mu \equiv e_2$, and the loop functions are as follows:
\begin{eqnarray}
F_\text{h}(x) &=& -\frac{ x^3 - 6 x^2 + 3 x  +  6 x \ln(x) + 2}   {6(1 - x)^4}, \\
G_\text{h}(x) &=& \frac{ x^2 - 4 x + 2  \ln (x) + 3 }{(1 - x)^3} .
\end{eqnarray}

The contribution from the $Z$ diagram is given by    
\begin{equation}
\delta a_\mu^Z = -\frac{m_\mu}{8 \pi^2 M_Z^2} \sum_{b=4,5}  \left[  (|g^{Z \mu e_b}_L|^2 + |g^{Z \mu e_b}_R|^2) \, m_\mu  F_Z(x_{Zb})   +  {\rm Re} \, ( g^{Z \mu e_b}_L g^{Z \mu e_b*}_R ) \, m_{e_b}  G_Z(x_{Zb})  \right] , \end{equation}
where $x_{Zb} = (m_{e_b}/M_Z)^2$, the couplings are given in Eqs.~(\ref{eq:gZL}) and (\ref{eq:gZR})  with index $\mu \equiv e_2$, and the loop functions are as follows:
\begin{eqnarray}
F_Z(x) &=& \frac{ 5 x^4 - 14x^3  + 39 x^2   - 38 x   - 18 x^2 \ln(x) + 8}{12  (1 - x)^4} ,  \\
G_Z(x) &=&  \frac{x^3  + 3 x  - 6 x \ln(x) - 4}   { 2 (1 - x)^3}.
\end{eqnarray}

Finally, the contribution from the $W$ diagram is given by  
\begin{equation}
\delta a_\mu^W = -\frac{m_\mu  }{16 \pi^2 M_W^2 }   \left[  ( |g^{W \nu_4 \mu}_L|^2 +|g^{W \nu_4 \mu}_R|^2) m_\mu  F_W(x_{W})   +  {\rm Re}\, ( g^{W \nu_4 \mu}_L g^{W \nu_4 \mu*}_R ) \, M_L G_W(x_{W})  \right],
\end{equation}
where $x_{W} = (M_L/M_W)^2$ since the mass of $\nu_4$ is given by $M_L$. The couplings are given in Eqs.~(\ref{eq:gWL}) and (\ref{eq:gWR})  with index $\mu \equiv e_2$, and the loop functions are as follows:
\begin{eqnarray}
F_\text{W}(x) &=& -\frac{ 4 x^4 - 49 x^3 + 78 x^2  - 43 x   + 18 x^3  \text{ln}(x) + 10 } { 6 (1 - x)^4} , \\
G_\text{W}(x) &=&  -\frac {x^3 - 12 x^2 + 15 x  +    6 x^2 \text{ln} (x)  -4   }  { (1 - x)^3 }.
\end{eqnarray}

 \subsection{Qualitative analysis}
 
An interesting insight can be obtained by integrating out vectorlike leptons~\cite{Kannike:2011ng}. 
In the limit (\ref{eq:integrateout})
the muon mass, after EW symmetry breaking,  receives contributions from two terms:
  \begin{equation}
 {{\cal L}_{eff}} \supset -  \bar \mu_{L}  \left( y_{\mu} + \frac{\lambda^L \bar \lambda \lambda^E }{M_L M_E} H H^\dag \right) \mu_{R} H  + h.c. \longrightarrow - \left( m^H_{\mu} + m_\mu^{LE} \right) \bar \mu_{L} \mu_{R} + h.c.,
 \label{eq:Leff}
 \end{equation}
 where $m^H_{\mu}$ originates from the direct Yukawa coupling of the muon flavor eigenstate and $m_\mu^{LE}$ comes from the mixing with vectorlike leptons. Due to the same chiral structure, the $m_\mu^{LE}$ term also contributes to the muon magnetic moment.  This contribution can be written as
  \begin{equation}
 \Delta a_\mu	\; \simeq \; c \, \frac{m_\mu m_\mu^{LE}}{(4 \pi v)^2}	\; \simeq \;  0.85 \, c \, \frac{m_\mu^{LE}}{m_\mu} \, \Delta a^{exp}_\mu.
 \label{eq:c}
 \end{equation}
 In the limit $M_E \simeq M_L \gg M_Z$ it was found that $c=-1$~\cite{Kannike:2011ng}. This means that the contributions to the muon mass and muon g-2 are anticorrelated. Nevertheless, ignoring the wrong sign,  the size of the contribution to the muon g-2 is what is needed to explain the measured value when  the muon mass originates mostly from the mixing with vectorlike leptons.

However, this conclusion holds only in the asymptotic region $M_E \simeq M_L \gg M_Z$. 
We can obtain a simple approximate formula for $\Delta a_\mu$ even in the region of small $M_E$ and $M_L$.  It turns out, that the formula (\ref{eq:c}) is still valid with $c$ being a function of masses of extra fermions which can be written as:
  \begin{equation}
c = c_W(x_W)  + c_Z (x_Z)+ c_h(x_h) \simeq   G_W(x_W) - 2 .
 \label{eq:cW}
 \end{equation}
 The second part  follows from $c_W(x_W) \simeq G_W(x_W)$ and the sum of the $c_Z (x_Z)$ and $c_h(x_h)$ being approximately $-2$ in a large range of masses, as a result of $G_Z(x)$ and  $ x G_h(x)$ changing slowly with $x$.
 
The Higgs contribution can be approximately written as $c_h(x_h) \simeq 3/2 \, x_h G_h(x_h) $ where $x_h$ is associated with  the lighter of the two leptons.\footnote{In the derivation of this formula we used the fact that $ x_hG_h(x_h)$ varies very little with $x_h$. A better approximation is $c_h(x_h) \simeq   x_{h4} G_h(x_{h4}) + 1/2 \, x_{h5} G_h(x_{h5})$ when the masses of the two charged leptons are different, and $c_h(x_h) \simeq  x_h G_h(x_h) - 1/2 x_h^2 G'(x_h)$ when the masses  are similar.} It varies from $-1$ to $-3/2$ for $x_h$ between $ 1$  and $ \infty$. Asymptotically, or if at least one of the masses  of extra charged leptons is somewhat larger than $M_Z$, the $Z$ contribution is given by $c_Z (x_Z) \simeq G_Z(x_Z) $,  where the $x_Z$ is associated with  the heavier of the two leptons. Numerically,
  $c_Z (x_Z) \simeq -1/2$ which is the asymptotic value of $G_Z(x_Z) $.
For both masses close to $M_Z$ we find $c_Z (x_Z) \simeq G_Z(x_Z) + x_ZG'(x_Z)$ which for $x_Z = 1$ equals $-3/4$. Therefore, the Z contribution varies from $-3/4$ to $-1/2$ for $x_Z$ between $ 1$  and $ \infty$. Thus,  the Z and Higgs contributions add up to $\sim -2$ for a large range of masses of vector like leptons. 
 
 The $W$ contribution, $c_W(x_W) \simeq G_W(x_W)$, strongly depends on the mass of the heavy neutrino, which in our model is given by $M_L$. While its asymptotic value is $+1$ leading to total $c \simeq -1$, for $M_L \simeq M_W$ it is  $+3$ leading to total $c \simeq +1$. Therefore, the correlation between  contributions to the muon mass and muon g-2 is mostly controlled by $M_L$, and we have two solutions: the asymptotic one,  $M_L \gg M_Z$, in which case the measured value of the muon g-2 is obtained for $ m_\mu^{LE}/m_\mu \simeq -1$; and the second one with a light extra neutrino, $M_L \simeq M_W$, in which case the measured value of the muon g-2 is obtain for $ m_\mu^{LE}/m_\mu \simeq +1$. In the first case, about twice as large contribution from the direct Yukawa coupling of the muon is required to generate the correct muon mass, while in the second case, the muon mass can fully originate from the mixing with heavy leptons. Any other correlation between $+1$ and $-1$ can be obtained by increasing the $M_L$ from the EW sale to  the $\sim 1$ TeV scale. 
 
\begin{figure}[t]
\includegraphics[width=3.in]{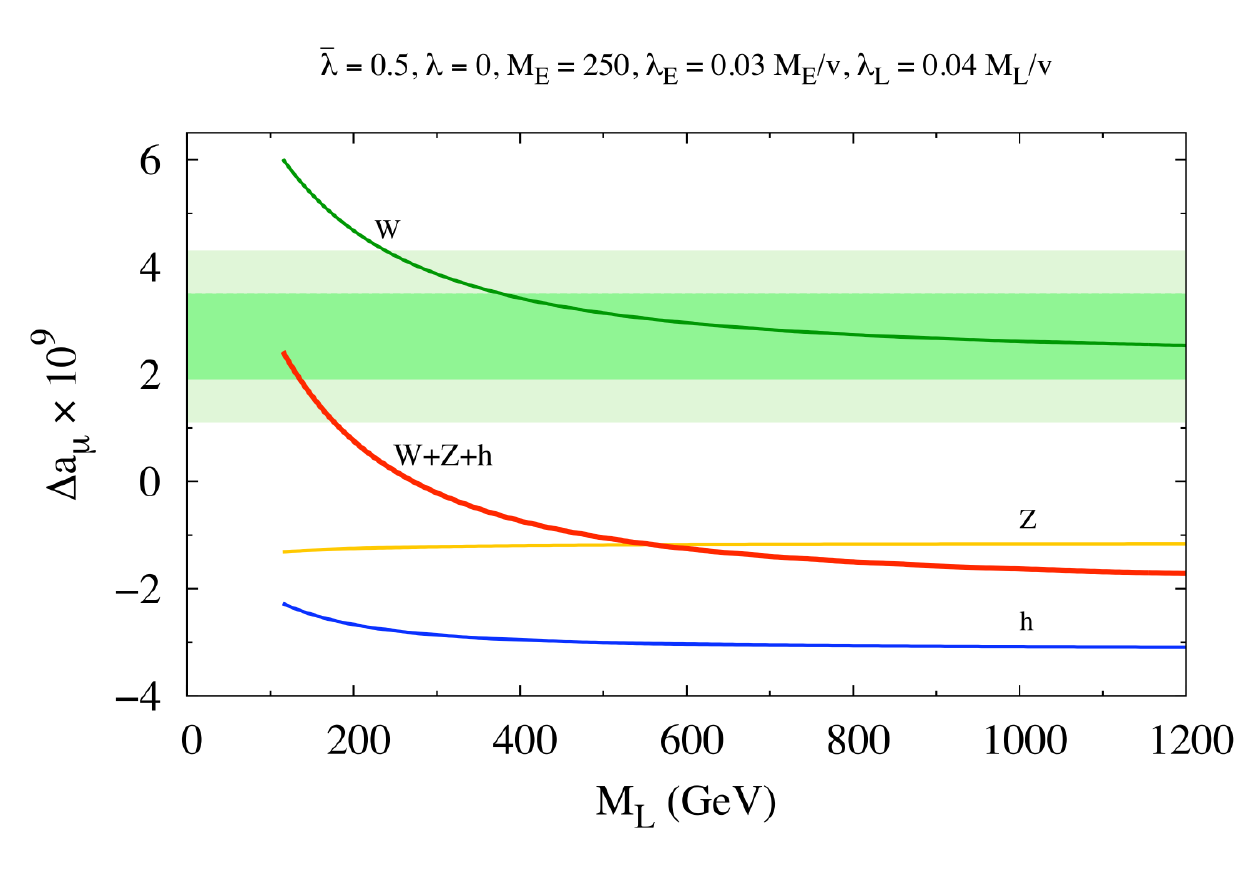} \hspace{0.5cm}
\includegraphics[width=3.in]{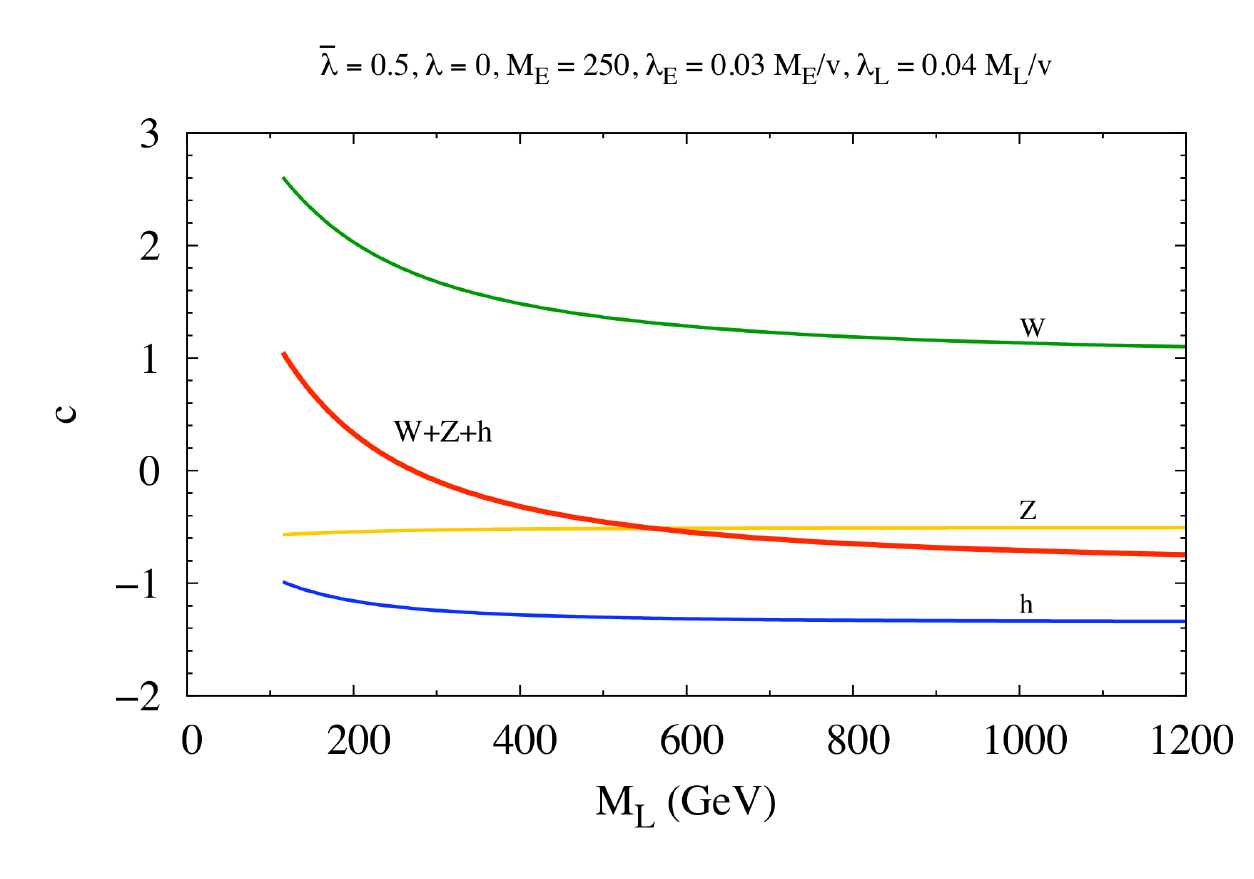}
\caption{Left:  contributions to the muon g-2 from $Z, W$ and $h$ loops with heavy leptons shown in Fig.~\ref{fig:g-2} as functions of $M_L$. The sum of all contributions is also plotted. Dark and light shaded bands correspond to  $1\sigma$ and $2\sigma$ regions of $\Delta a_\mu$  specified in Eq.~(\ref{eq:del_a_mu}).  Right: separate and total contributions to the $c$ coefficient defined by assuming the equality in Eq.~(\ref{eq:c}) that show the correlation between contributions of heavy leptons to the muon g-2 and the  muon mass. In both plots we fix: $M_E = 250 $ GeV, $\bar \lambda = 0.5$, $ \lambda = 0$ and $\lambda^L$ and $\lambda^E$ are set to their approximate maximum allowed values given in Eq.~(\ref{eq:lambda_ELmax}). The signs of couplings are chosen so that $m^{LE}_\mu$ is positive. For the opposite  sign of $m^{LE}_\mu$, the signs on the y-axis in the left plot should be flipped.}
\label{fig:contributions}
\end{figure}

 This is illustrated in Fig.~\ref{fig:contributions} in which we show separate contributions to the muon g-2 and the $c$ coefficient from $Z, W$ and $h$ as functions of $M_L$. In both plots we fix: $M_E = 250 $ GeV, $\bar \lambda = 0.5$, $ \lambda = 0$, and $\lambda_L$ and $\lambda_E$ are set to their approximate maximum values allowed by precision EW data (discussed later) which fixes $m^{LE}_\mu$. Almost identical shape of lines in both plots in Fig.~\ref{fig:contributions} further supports the fact that Eq.~(\ref{eq:c}) with $c$ given in (\ref{eq:cW}) is indeed a good approximation. Chosen  signs of couplings in the plots correspond to generating a positive contribution to the muon mass from $m^{LE}_\mu$. If the product  $\lambda^L \bar \lambda \lambda^E$ is negative, leading to a negative contribution to the muon mass from $m^{LE}_\mu$,  the g-2 plot would look identical with the signs on the y-axis flipped. 
 
The contributions to the $c$ coefficient in  Fig.~\ref{fig:contributions}, for a given choice of parameters, are representative for a large range of $M_L$ and $M_E$. The plots would  be almost identical for any larger value of $M_E$, and would only slightly change in the small $M_L$ region for $M_E$ as small as 100 GeV. All the Yukawa couplings only rescale the contributions to the muon g-2, different choices do not change the results qualitatively as far as the condition (\ref{eq:integrateout}) is satisfied.

\subsection{Constraints}

In the numerical scans over the parameter space that follow,
we impose constraints from precision EW data related to the muon that include the $Z$ pole observables ($Z$ partial width, forward-backward asymmetry, left-right asymmetry), the $W$  partial width, and the muon lifetime~\cite{PDG}. 
In the limit of small couplings (\ref{eq:integrateout}) these constraints approximately translate into 95\% C.L. bounds on $\lambda^{E,L}$ couplings:
  \begin{equation}
\frac{\lambda^E v}{M_E} \lesssim 0.03, \quad \quad \frac{\lambda^L v}{M_L} \lesssim 0.04.
 \label{eq:lambda_ELmax}
 \end{equation}
 These quantities squared represent modifications of the SM couplings of the $Z$ and $W$  to the muon, which can be obtained from Eqs.~(\ref{eq:delgL}), (\ref{eq:delgR}), (\ref{eq:gWL}) and the diagonalization matrices (\ref{eq:UL}) and (\ref{eq:UR}).
We further impose constraints from oblique corrections, namely from S and T parameters~\cite{PDG}.
Finally, we impose the LEP limits on masses of  charged leptons which are required to be larger than 105 GeV.

 \subsection{Scan over the parameter space: the muon g-2 and Higgs decays}
 
\begin{figure}[t]
\includegraphics[width=3.in]{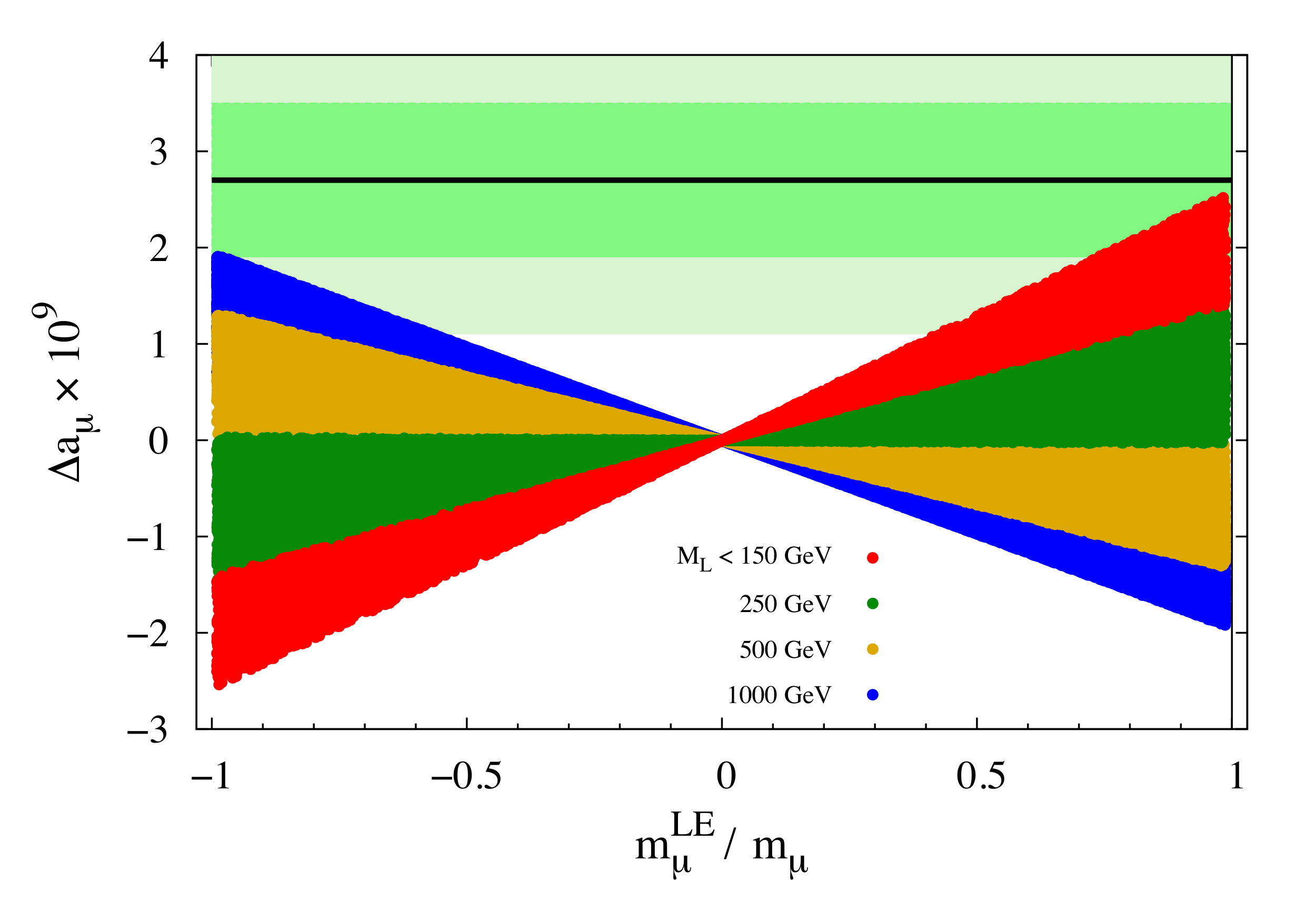} \hspace{0.5cm}
\includegraphics[width=3.in]{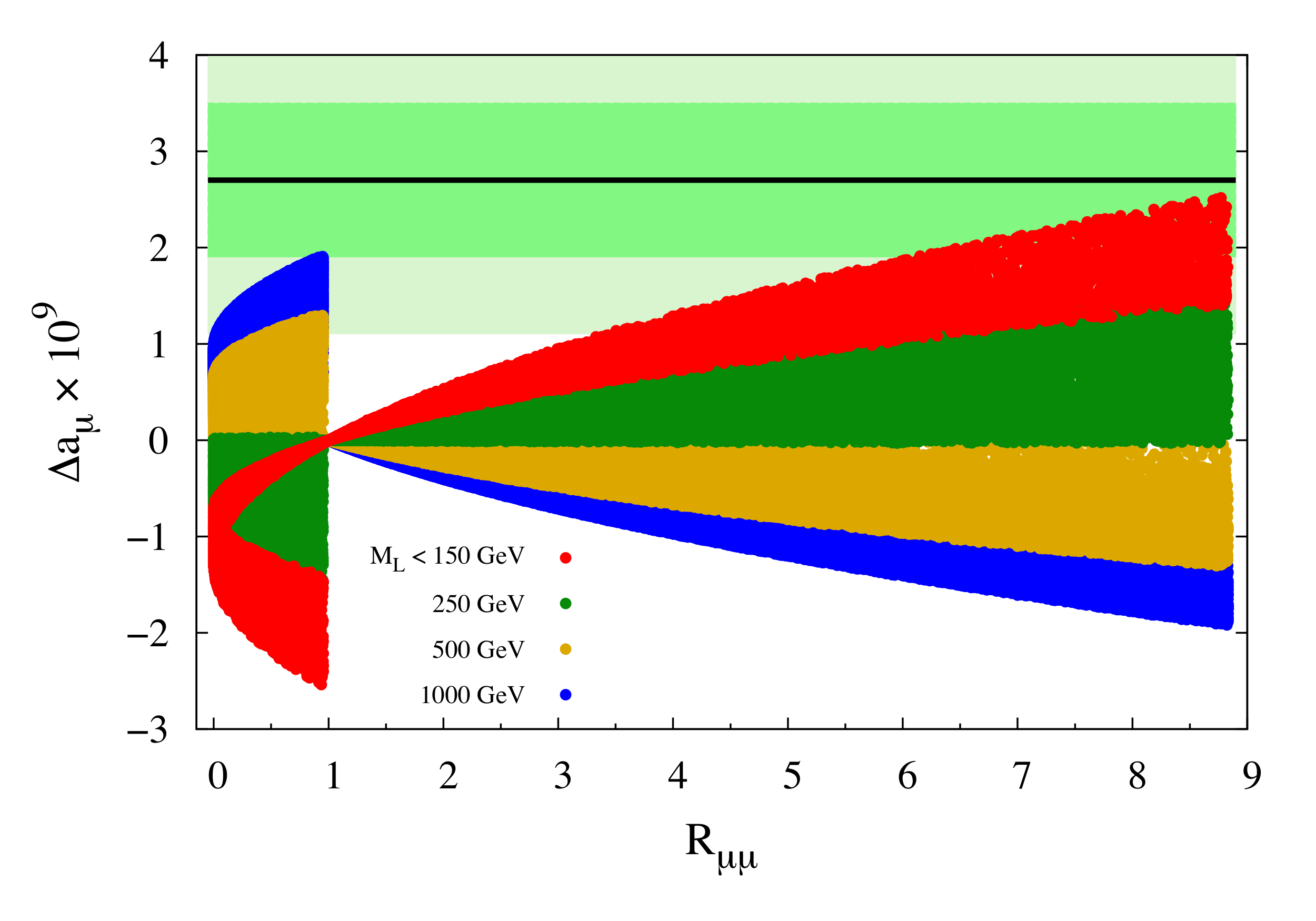}
\caption{Left: Randomly generated points with $M_L \in [100, 1000]$ GeV, $M_E \in [100, 1000]$ GeV, $\bar \lambda < 0.5$, $\lambda = 0$, and $\lambda_{L,E}$  in allowed ranges from precision EW data, plotted in $\Delta a_\mu$ -- $m_\mu^{LE}/m_\mu$ plane. Both signs of couplings are allowed. The lightest mass eigenstate is required to satisfy the LEP limit. Different colors (shades) correspond to different regions of $M_L$ in the order from top to bottom on the right side of the plot: $M_L < 150$ GeV, $< 250$ GeV, $< 500$ GeV, and $< 1000$ GeV (the order is reversed on the left side of the plot).
Horizontal line and dark (light) shaded bands correspond to the central experimental value of $\Delta a_\mu$ and $1\sigma$ ($2\sigma$) regions respectively, specified in Eq.~(\ref{eq:del_a_mu}).
Right: the same points as on the left plotted in $\Delta a_\mu$ -- $R_{\mu\mu}$ plane.
 }
\label{fig:p5p0}
\end{figure}

 Previous qualitative discussion of expected results is fully supported by numerical scans. In Fig.~\ref{fig:p5p0} on the left we plot the contribution to the muon g-2 versus the contribution to the muon mass from the mixing with heavy leptons for randomly generated points with $M_L \in [100, 1000]$ GeV, $M_E \in [100, 1000]$ GeV, $\bar \lambda < 0.5$, and $\lambda_{L,E}$  in allowed ranges from precision EW data. For simplicity,  $\lambda$ is set to 0 in these plots, because it should not have a significant effect on our results.  In all the plots in this section the $m^{LE}_\mu$ is defined more precisely as the mass that the muon would have if the direct Yukawa coupling was zero. Different colors (shades) correspond to different regions of $M_L$. This shows   that it is indeed $M_L$ that controls the correlation between the contribution to the muon g-2 and muon mass. 
 
There are two solutions: the asymptotic solution for large $M_L$ in which the measured muon g-2 can be obtained for $m_\mu^{LE}/m_\mu \simeq -1$ and so the physical muon mass is a result of a cancellation between the direct Yukawa coupling and the contribution from the mixing; and the light neutrino solution for $M_L \simeq 100$ GeV in which case the muon mass can fully originate from the mixing, $m_\mu^{LE}/m_\mu \simeq +1$.

On the right in Fig.~\ref{fig:p5p0} we plot the same points in $\Delta a_\mu$ -- $R_{\mu\mu}$ plane, where
  \begin{equation}
 R_{\mu\mu} \equiv \frac{\Gamma(h \to \mu^+ \mu^-)}{\Gamma(h \to \mu^+ \mu^-)_{SM}}.
 \label{eq:Rmumu}
 \end{equation}
This plot can be easily understood from Eq.~(\ref{eq:Leff}). The enhancement of $\Gamma(h \to \mu^+ \mu^-)$ by a factor of 9 compared to the SM in the small $M_L$ case that can explain the muon g-2 anomaly, for which $m_\mu^{LE}/m_\mu \simeq +1$,  originates  from 3 possible ways one Higgs coupling and 2 vevs replace the Higgs fields in the second term in Eq.~(\ref{eq:Leff}). In the asymptotic case, $m_\mu^{LE}/m_\mu \simeq -1$, that also explains the anomaly,  $R_{\mu\mu} \simeq 1$, which results from the first term in Eq.~(\ref{eq:Leff}) being twice as large as in the SM and the same combinatoric factor of 3 with a minus sign from the second term.

  \begin{figure}[t]
\includegraphics[width=3.in]{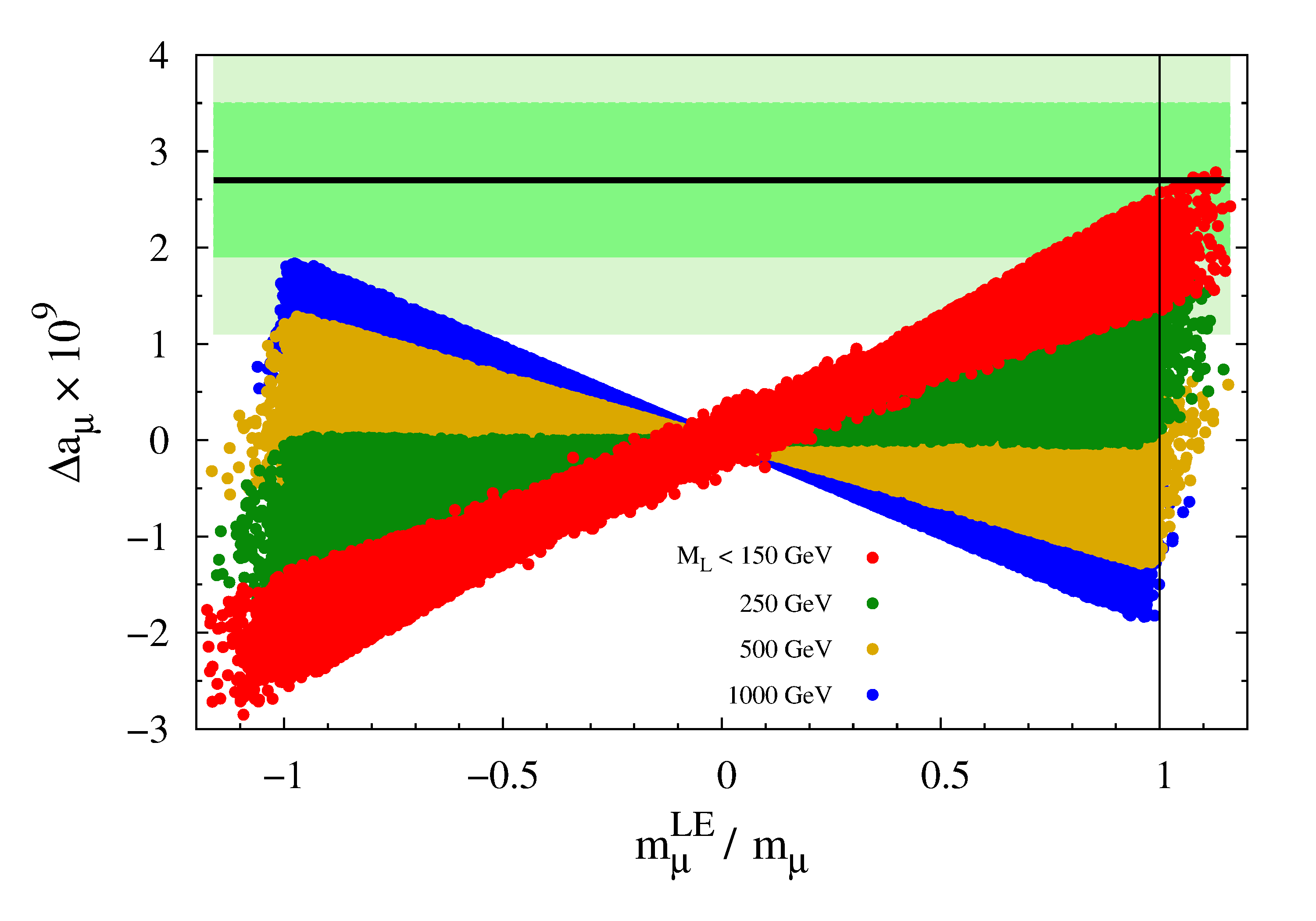} \hspace{0.5cm}
\includegraphics[width=3.in]{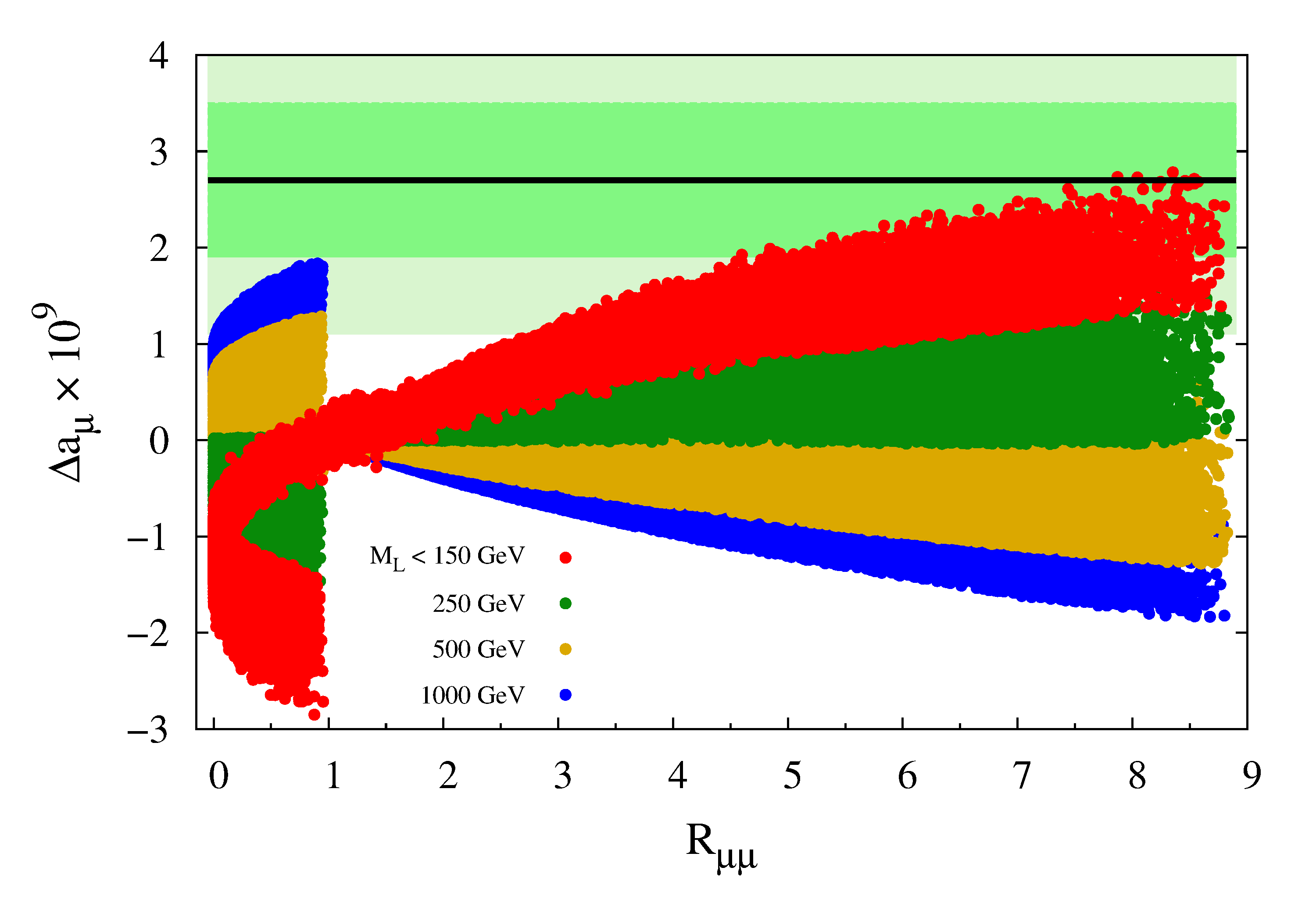}
\caption{The same as in Fig.~\ref{fig:p5p0} with additional randomly generated $ \lambda < 0.5$.}
\label{fig:p5p5}
\end{figure}
 
\begin{figure}[t]
\includegraphics[width=3.in] {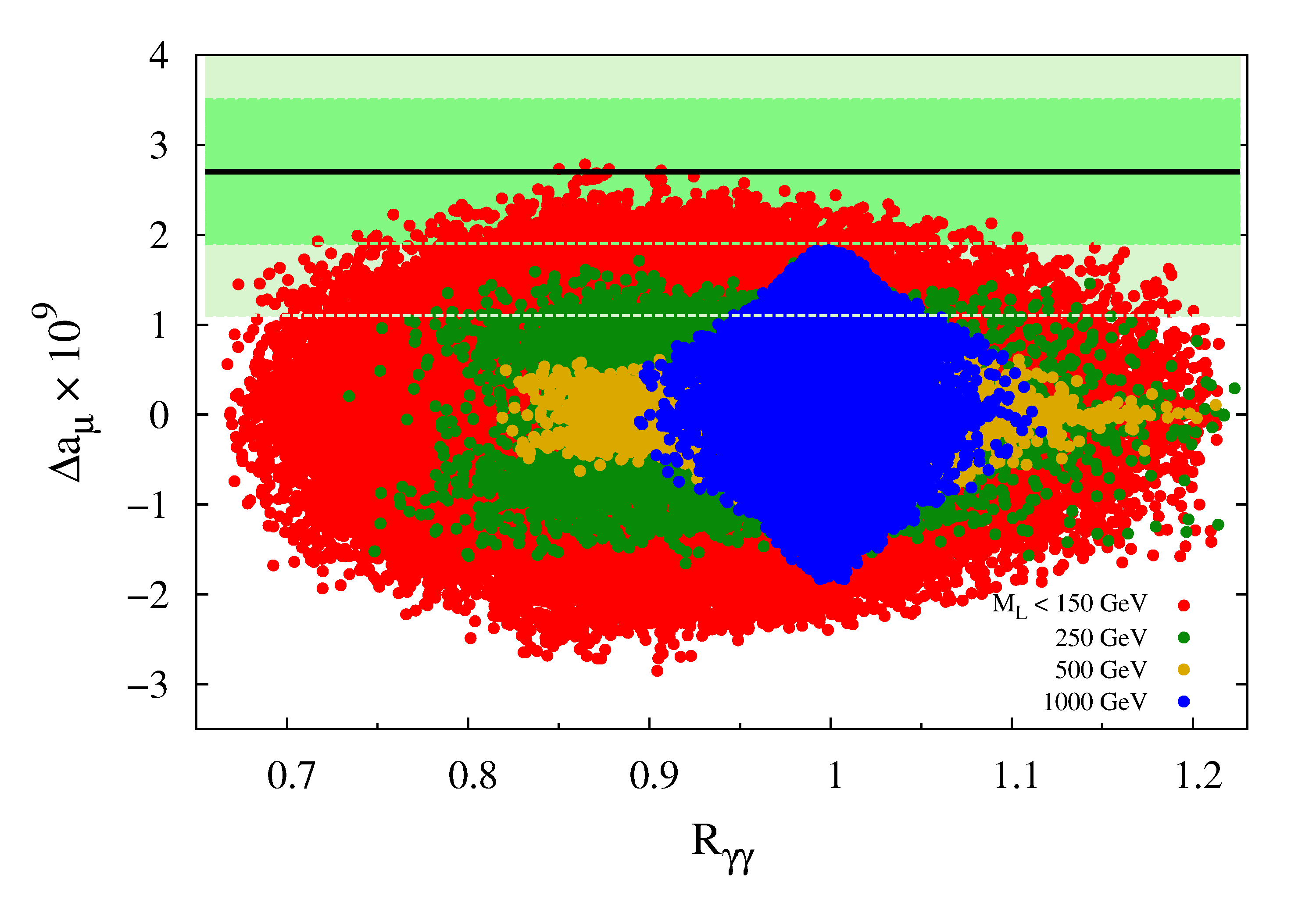}\hspace{0.5cm}
\includegraphics[width=3.in]{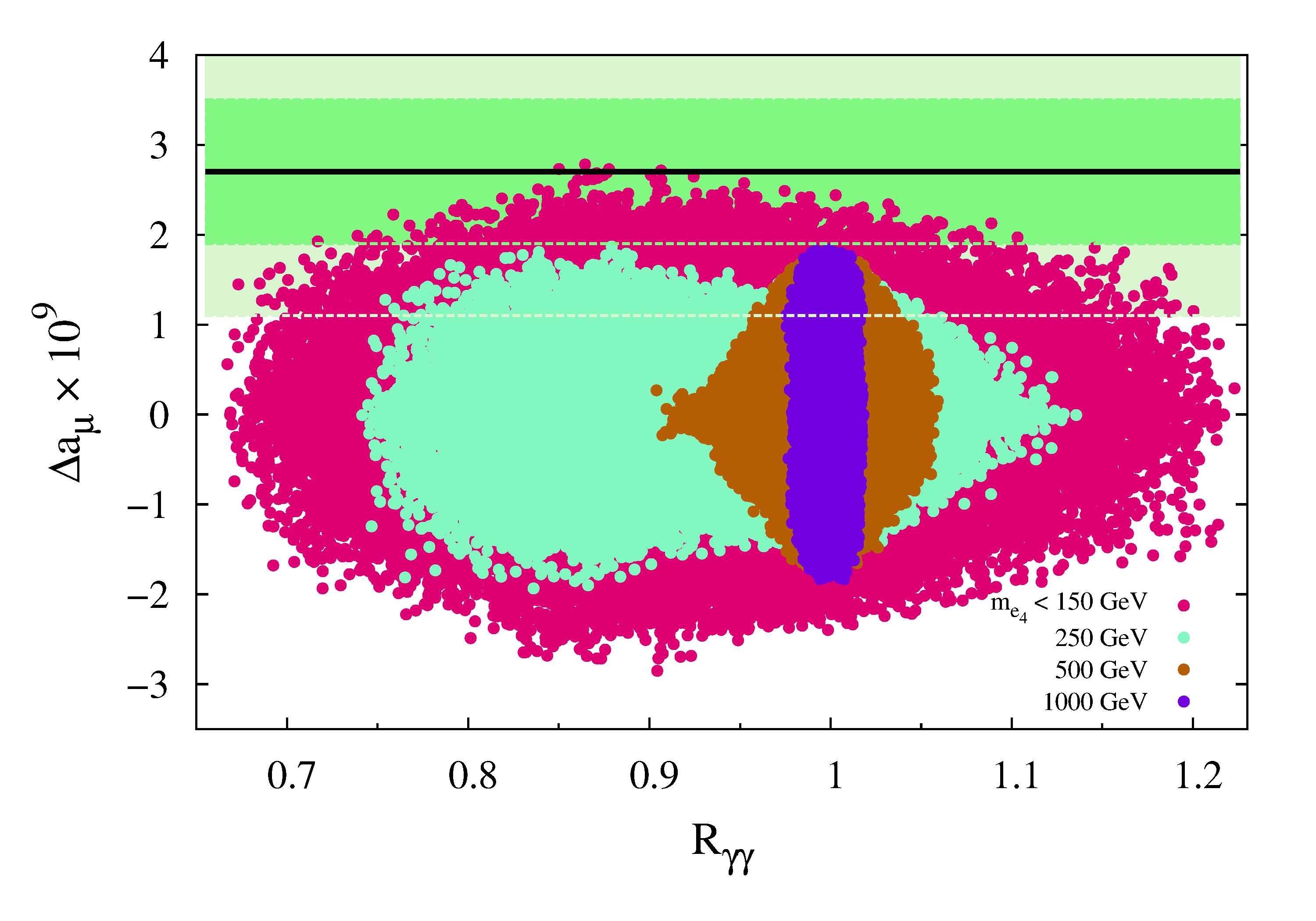}
\caption{Left: the same as in Fig.~\ref{fig:p5p5} plotted in the $\Delta a_\mu$ -- $R_{\gamma \gamma}$ plane. Right: the same as on the left but different colors (shades) represent the mass of the lighter extra charged lepton mass eigenstate, $m_{e_4}$, in ranges  $< 150$ GeV, $< 250$ GeV, $< 500$ GeV, and $< 1000$ GeV when going from outside toward the center. }
\label{fig:p5p5HGG}
\end{figure}

From the qualitative discussion in the previous section we expect that allowing nonzero $\lambda$ should not change the results dramatically. This can be seen in Fig.~\ref{fig:p5p5} which is obtained under the same conditions as Fig.~\ref{fig:p5p0} with additional randomly generated $\lambda < 0.5$. Additional $\lambda$ coupling has however important consequences for $h \to \gamma \gamma$.  If both $\lambda$ and $\bar \lambda$ are nonzero, the $h \to \gamma \gamma$ can be significantly modified. In Fig.~\ref{fig:p5p5HGG} we plot  the points from Fig.~\ref{fig:p5p5} in the $\Delta a_\mu$ -- $R_{\gamma \gamma}$ plane, where
    \begin{equation}
 R_{\gamma\gamma} = \frac{\Gamma(h \to \gamma \gamma)}{\Gamma(h \to \gamma \gamma)_{SM}}.
 \label{eq:Rgammagamma}
 \end{equation}
In the plot on the left the color notation is the same as in Fig.~\ref{fig:p5p5}, namely it represent different regions of $M_L$, while in the plot on the right different colors (shades) represent the mass of the lightest mass eigenstate which is more meaningful for $h \to \gamma \gamma$.

In the small $M_L$ case that can explain the muon g-2 within one standard deviation the $R_{\gamma\gamma} $ can be decreased by about 25\% or increased by about 15\%. In the  asymptotic case, $R_{\gamma\gamma} $ is negligibly modified in the region that explains the muon g-2 within one standard deviation.

Some of the results presented so far depend on our upper limit on possible Yukawa couplings which we took to be 0.5. This upper limit is motivated by a simple UV embedding with nice features concerning the stability of the EW minimum of the Higgs potential, that we will discuss in the next section. This however is just one possibility and in principle larger values of Yukawa couplings should be considered on phenomenological grounds. Thus we also plot similar results assuming upper limit for all Yukawa couplings to be 1.

\begin{figure}[t]
\includegraphics[width=3.in]{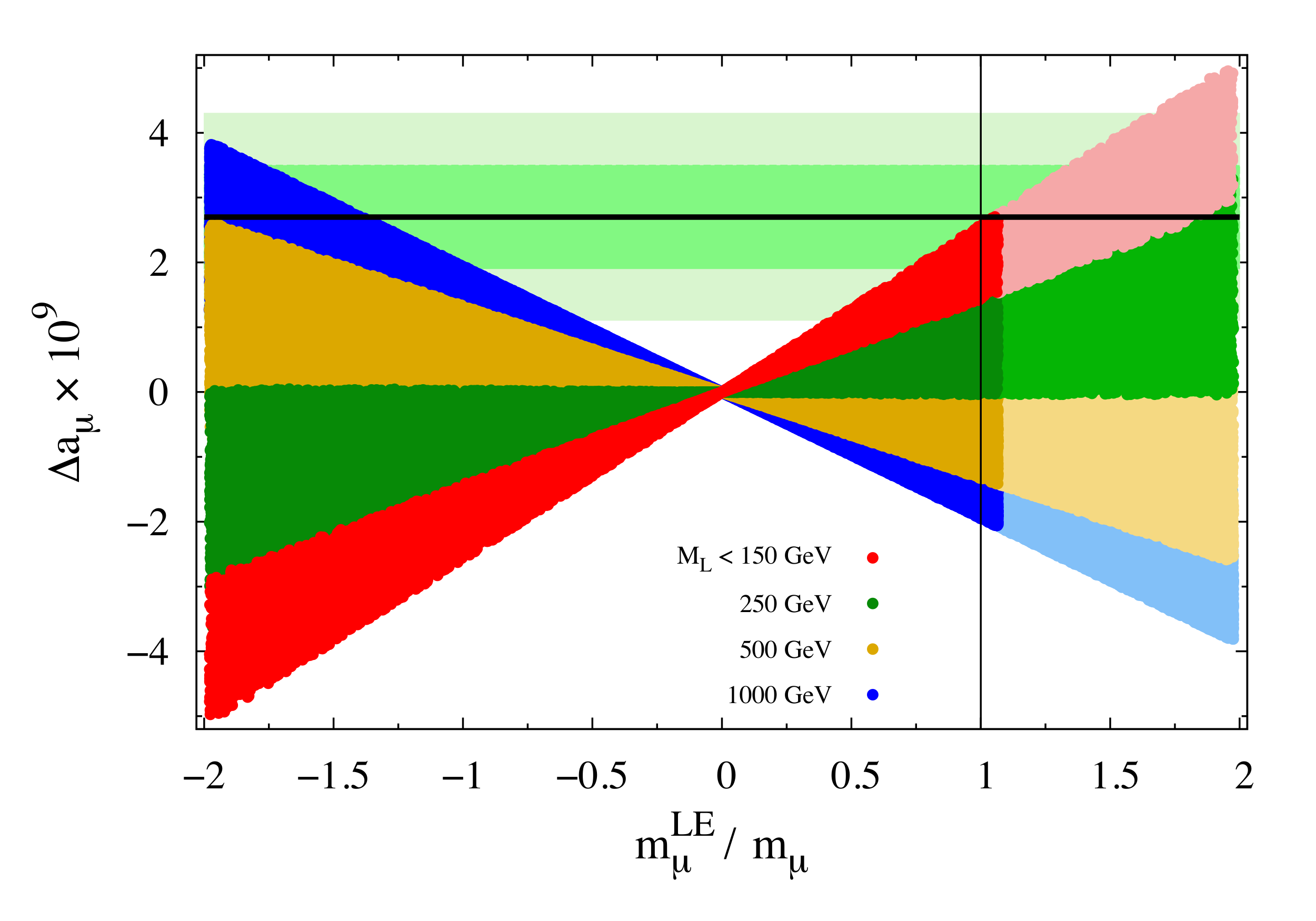} \hspace{0.5cm}
\includegraphics[width=3.in]{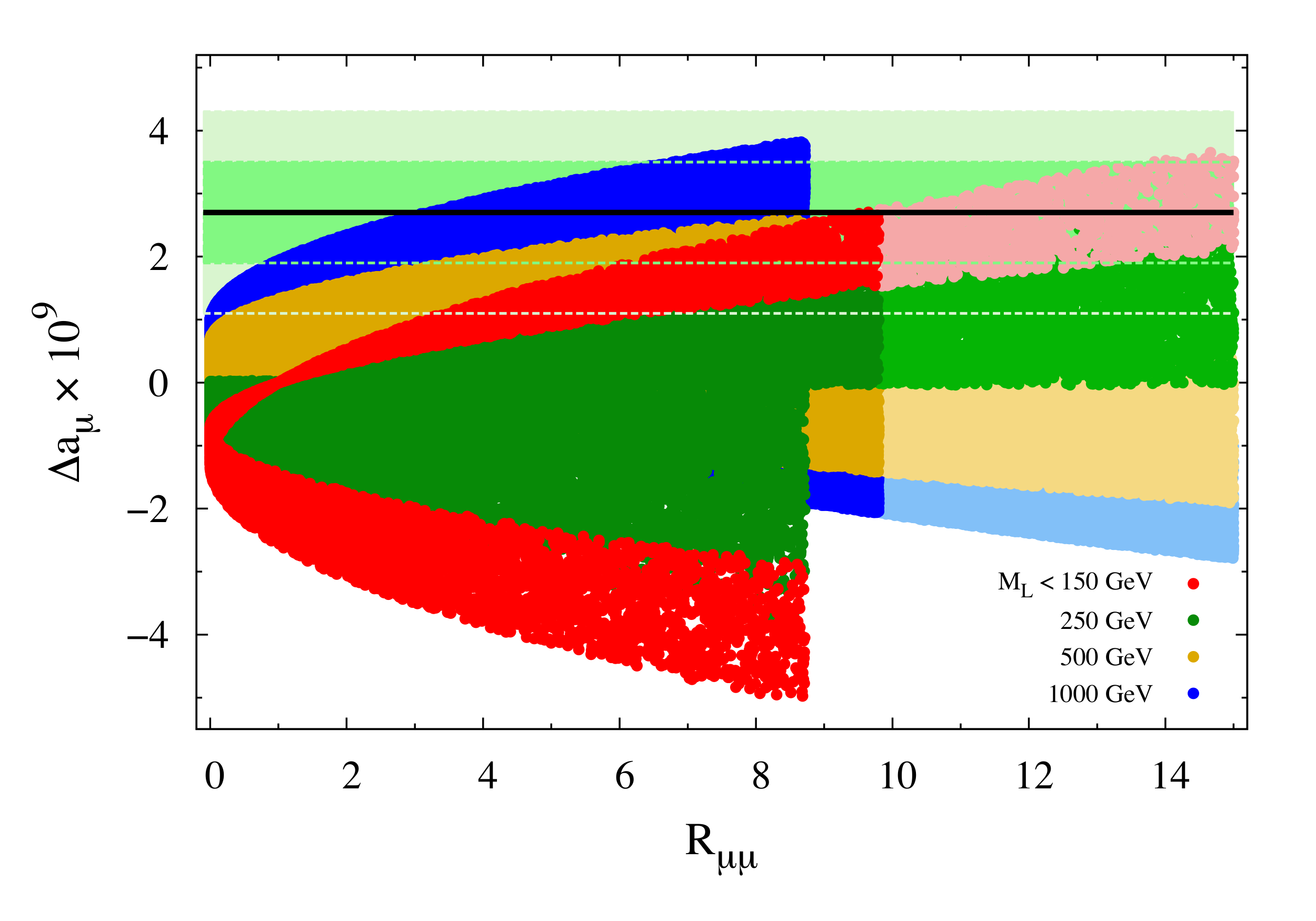}
\caption{The same as in Fig.~\ref{fig:p5p0} with the region of $ \bar \lambda$ extended to $ \bar \lambda < 1$. The lightly shaded points are excluded by the ATLAS search for $h\to \mu^+ \mu^-$. The plot on the right would extend to larger values of $R_{\mu\mu} \simeq 24$. }
\label{fig:1pp0}
\end{figure}

The randomly generated points extended to  $\bar \lambda < 1$ with $ \lambda = 0$ are plotted in Fig.~\ref{fig:1pp0} in the $\Delta a_\mu$ -- $m_\mu^{LE}/m_\mu$ plane on the left, and in  the $\Delta a_\mu$ -- $R_{\mu\mu}$ plane on the right. Comparing Figs.~\ref{fig:p5p0} and \ref{fig:1pp0} clearly demonstrates the effect of varying $\bar \lambda$ coupling; increasing $\bar \lambda$ extends the plots to larger values of both $\Delta a_\mu$ and $m_\mu^{LE}/m_\mu$, while the correlation of these contributions is unchanged. The Higgs coupling to the muon is modified more dramatically, and part of the parameters space is already ruled out by the ATLAS search for $h \to \mu^+ \mu^-$~\cite{Atlas_h_to_mumu} (lightly shaded regions in both plots). The  $R_{\mu\mu}$ ranges between 6 and 9.8 (the current limit) in the small $M_L$ case that can explain the muon g-2 anomaly, and between 1 and 9 in the asymptotic case.

\begin{figure}[t]
\includegraphics[width=3.in]{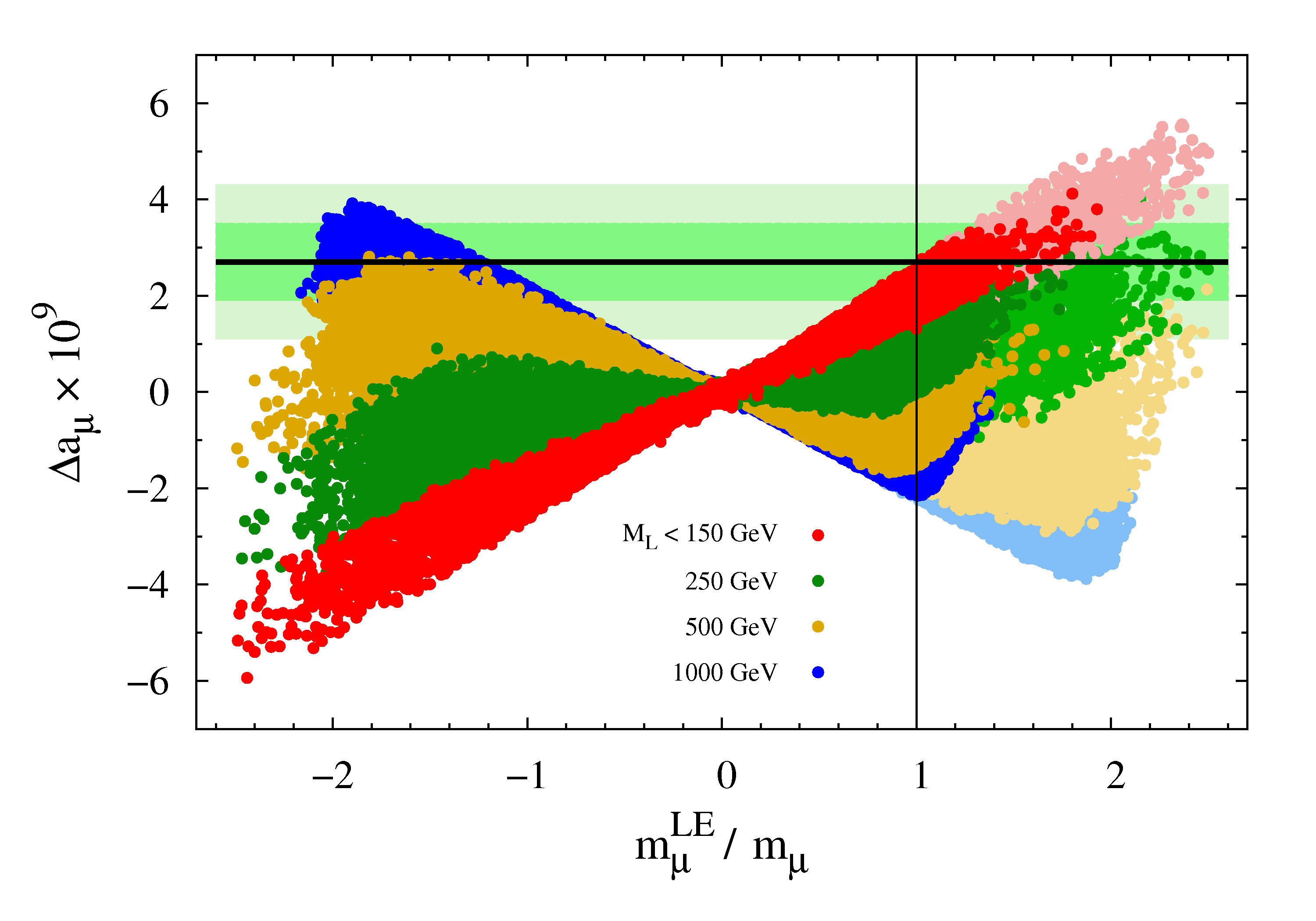} \hspace{0.5cm}
\includegraphics[width=3.in]{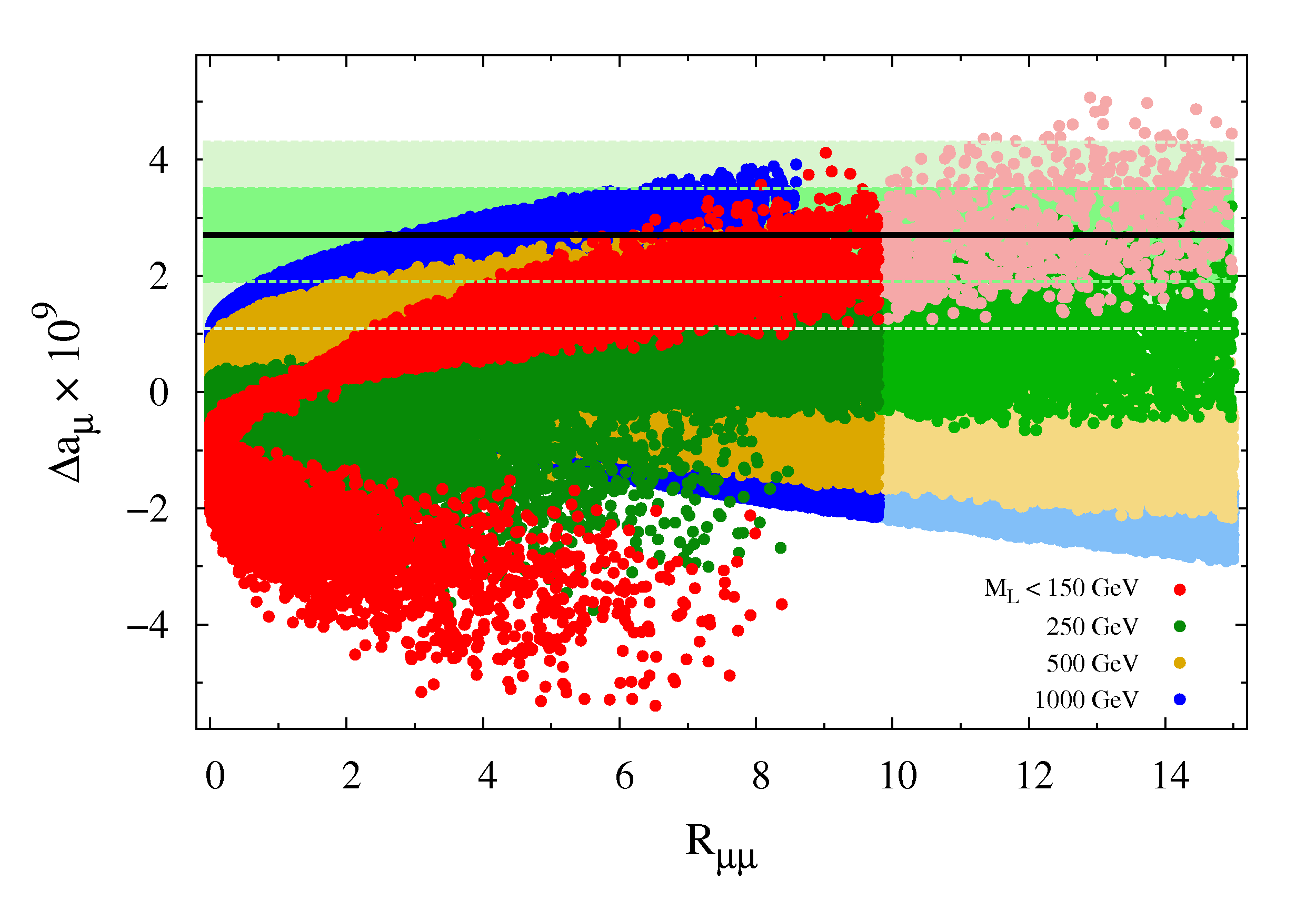}
\caption{The same as in Fig.~\ref{fig:1pp0} with additional randomly generated $ \lambda < 1$.}
\label{fig:1p1p}
\end{figure}

The addition of $\lambda <1$ somewhat expands the ranges of $\Delta a_\mu$, $m_\mu^{LE}/m_\mu$ and $R_{\mu\mu}$ for all regions of $M_L$, which can be seen in Fig.~\ref{fig:1p1p}. Nevertheless, as expected, the plots in Figs.~\ref{fig:1pp0} and \ref{fig:1p1p} look qualitatively very similar. 

\begin{figure}[t]
\includegraphics[width=3.in] {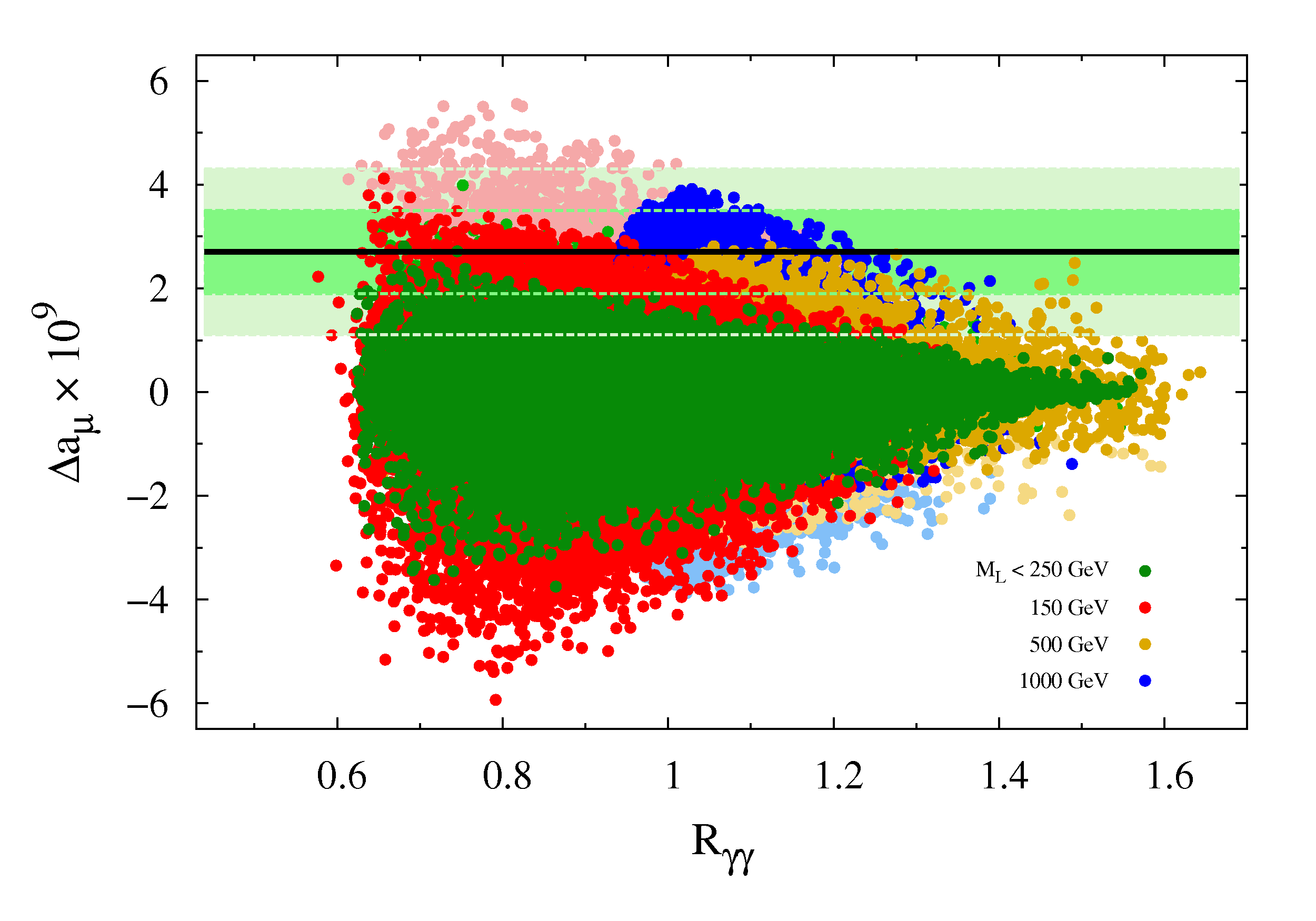}\hspace{0.5cm}
\includegraphics[width=3.in]{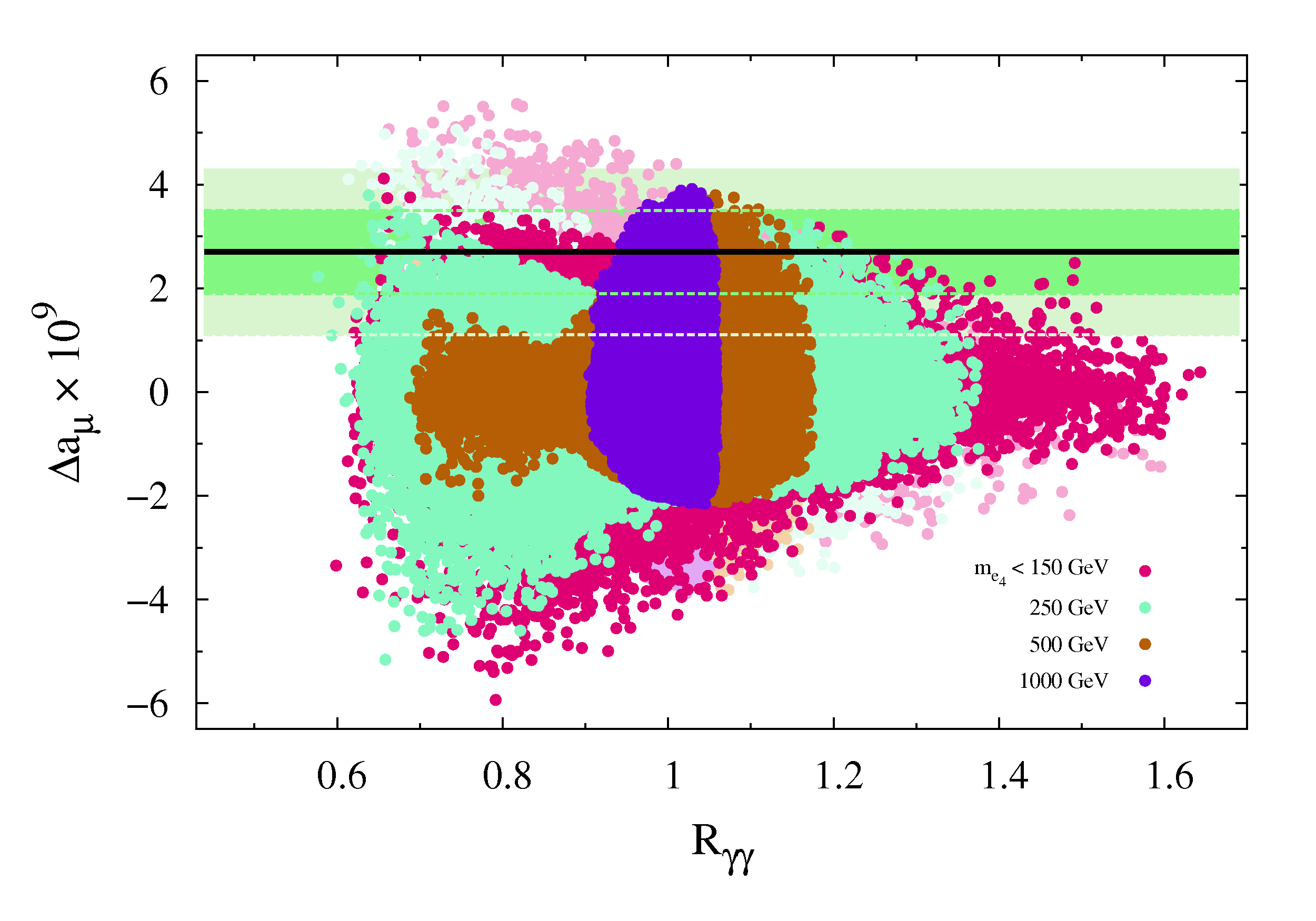} 
\caption{Left: the same as in Fig.~\ref{fig:p5p5} plotted in the $\Delta a_\mu$ -- $R_{\gamma \gamma}$ plane. Right: the same as on the left but different colors (shades) represent the mass of the lighter extra charged lepton mass eigenstate, $m_{e_4}$, in ranges  $< 150$ GeV, $< 250$ GeV, $< 500$ GeV, and $< 1000$ GeV when going from outside toward the center. The lightly shaded points are excluded by the ATLAS search for $h\to \mu^+ \mu^-$.}
\label{fig:1p1pHGG}
\end{figure}

Increasing $\bar \lambda$ and $ \lambda$ up to 1 significantly extends ranges of predictions for  $R_{\gamma \gamma}$, given in Fig.~\ref{fig:1p1pHGG},  especially for the asymptotic case. In the small $M_L$ case that can explain the muon g-2 anomaly within $1\sigma$, the $R_{\gamma \gamma}$ ranges between 0.6 and and 1.15; while in the asymptotic case the $R_{\gamma \gamma}$ ranges between 0.9 and 1.5. In addition, the range of possible masses of the lightest extra charged lepton significantly expands, see Fig.~\ref{fig:1p1pHGG} on the right. With $\bar \lambda, \lambda < 1$ for the small $M_L$ case that can explain the muon g-2 anomaly within $1\sigma$ the mass of the lightest extra charged lepton, $m_{e_4}$,  has to be at most $\sim$250 GeV, while with $\bar \lambda, \lambda < 0.5$ the maximum mass is only $\sim$150 GeV.

 \subsection{Light charged leptons at the LHC}
 
 This scenario, especially the small $M_L$ case  that can explain the muon g-2 anomaly  could be searched for at the LHC. The LHC phenomenology of extra leptons  was discussed for example in Refs.~\cite{Giudice:2008uua, DelNobile:2009st, Martin:2009bg},~\cite{Kannike:2011ng}.
 The pair production cross section of extra leptons with masses of order 100 GeV is about 1pb  at the LHC at 8 TeV (and it is steeply falling with increasing the mass). Extra leptons decay into $Z, W$ or $h$ and a light lepton. The decay branching ratios are typically comparable and so the signatures of this scenario are spread over a variety of final states. Especially for small masses of charged leptons the branching ratios highly depend on the Yukawa couplings. 
 
 So far limited searches have been done. A search motivated by heavy leptons specific to Type III Seesaw models limits $\sigma(pp \to L^0 L^\pm) \times B(L^\pm \to Z l^\pm) \times B(L^0 \to W l)$, where $l$ is either $e$ or $\mu$,  to about 200 fb for heavy leptons with the mass 100 GeV~\cite{Atlas_seesawIII}. There is also a similar search at  CMS for both heavy leptons decaying through $W$~\cite{CMS:2012ra}. In addition to these specific searches there are also general searches for  anomalous production of multi lepton events~\cite{Chatrchyan:2012mea} that constrain specific decay modes of heavy leptons.
 
However, in addition to the dependance of the branching ratios on Yukawa couplings within the scenario we discussed,  these in principle also  depend on couplings that are not necessary for the explanation of the muon g-2. For example,  heavy leptons can dominantly decay into $\tau$ leptons reducing the number of light leptons in final states. Due to limited existing constraints, all the scenarios we discuss are or easily can be made viable. This could dramatically change with further searches for heavy leptons covering all the decay modes in near future. In addition, improving the limits on  $h\to \mu^+ \mu^-$, that already constrain parts of the parameter space, is highly motivated even if the sensitivity to the SM prediction for this process cannot be reached soon. 

Finally, the charged leptons relevant to the asymptotic solution are  far beyond the current reach of the LHC. However, this solution can still  be highly constrained by improving the limits on $h\to \mu^+ \mu^-$.

 \section{A possible UV completion}
  \label{sec:SM+3VFs}
 
 The model with extra vectorlike leptons can be embedded into recently discussed scenario with extra 3 or more complete  vectorlike families~\cite{Dermisek:2012as, Dermisek:2012ke}. 
 This scenario features gauge coupling unification, sufficiently stable proton, and the Higgs quartic coupling remaining positive all the way to the GUT scale. Predicted values of gauge couplings at the electroweak scale are highly insensitive to GUT scale parameters and masses of vectorlike fermions. They can be understood from  IR fixed point predictions and threshold effects from integrating out vectorlike families.
 
\begin{figure}[t]
\includegraphics[width=2.8in]{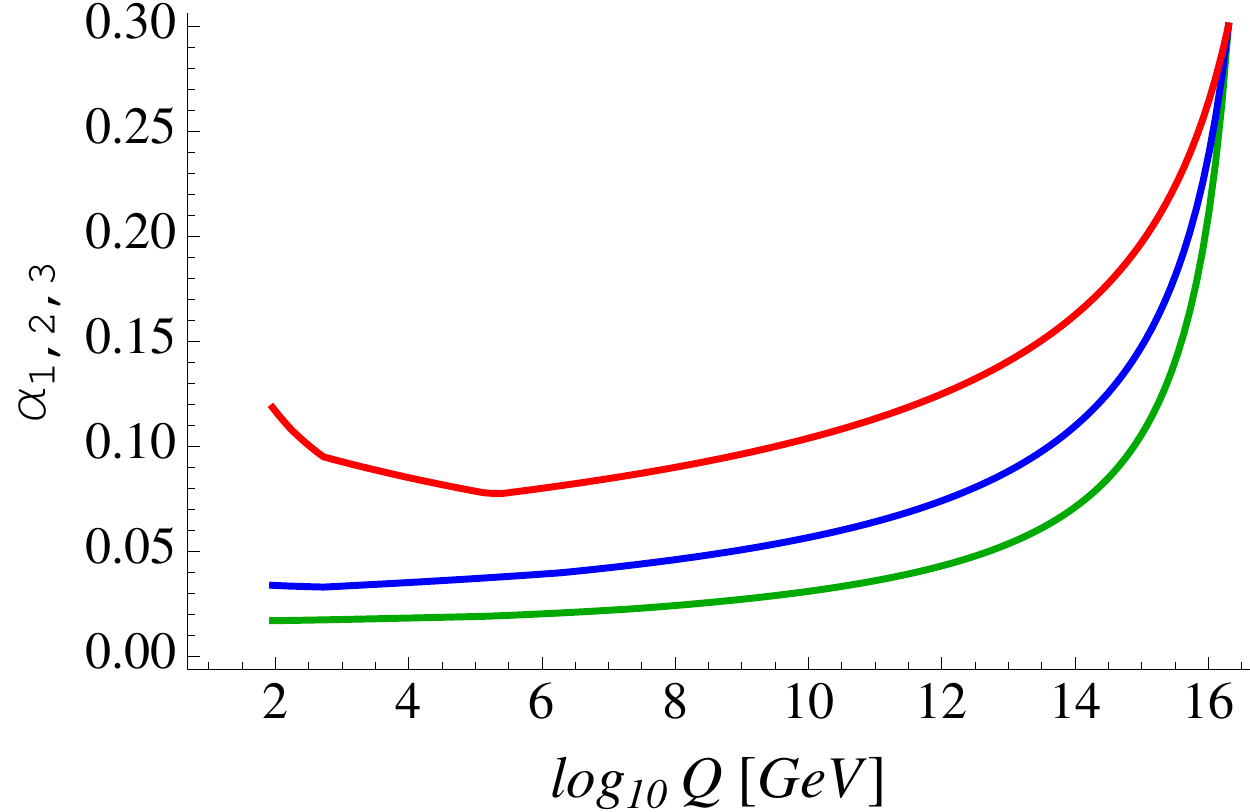} \hspace{0.5cm}
\includegraphics[width=2.8in]{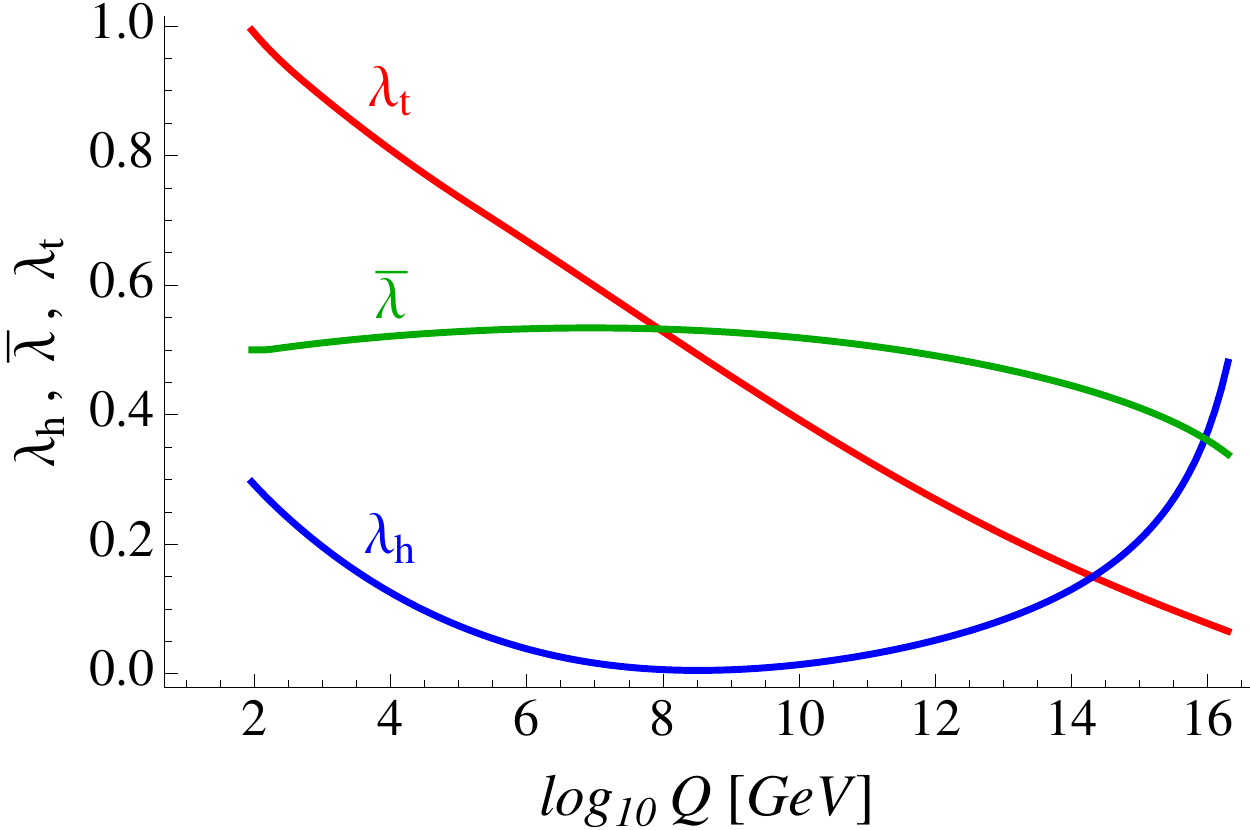}
\caption{Left: the RG evolution of gauge couplings: $\alpha_3$ (top), $\alpha_2$ (middle), and $\alpha_1$ (bottom) in the SM extended by three vector-like families for $\alpha_G = 0.3$ at $M_G = 2 \times 10^{16}$ GeV. Right: the RG evolution of the Higgs quartic coupling for $m_h = 126$ GeV, the top Yukawa coupling and the $\bar \lambda$ with the EW scale value of 0.5. Masses of vector like fermions are $M_{L_1} = M_{E_1}=150$ GeV, $M_{L_{2,3}} = 2.0 \times 10^6$ GeV, $M_{E_{2,3}} = 2.4 \times 10^7$ GeV, $M_{Q} = 520$ GeV, $M_{U} = 1.4 \times 10^5$ GeV, and $M_{D} = 2.5  \times 10^5$ GeV.}
\label{fig:RG}
\end{figure}

 These features are preserved even  when one generation of L and E have masses close to the EW scale and the extra Yukawa couplings are of the size required to obtain the measured value of the muon magnetic moment. A specific example assuming extra three complete vectorlike familes is given in Fig.~\ref{fig:RG}. We fix $M_{L_1}$ and $M_{E_1}$ to 150 GeV (even sizable variations of these masses would have negligible impact on the results presented in this section) and $\bar \lambda$ to 0.5. This value of $\bar \lambda$ and masses of vectorlike leptons can generate muon g-2 close to the central value and simultaneously generate the muon mass completely from the mixing between light and heavy families. The masses of the other two generations of vectorlike leptons and all three generations of quarks are varied to obtain the measured values of gauge couplings at the EW scale starting from $\alpha_G = 0.3$ at $M_G = 2 \times 10^{16}$ GeV. We set all other Yukawa couplings to zero, except for the top quark Yukawa. The contributions from $\lambda^L$ and $\lambda^E$ to the RG evolution of gauge, top Yukawa and Higgs quartic couplings  can also be neglected when the constraints from precision EW data are satisfied. The analysis closely follows~\cite{Dermisek:2012as, Dermisek:2012ke} with the only exception that  we use 1-loop  RGEs for Yukawa and Higgs quartic couplings.   We use 2-loop RGEs for gauge couplings as in Ref.~\cite{Dermisek:2012as, Dermisek:2012ke}.
 
 The evolution of gauge couplings is almost identical to examples given  in Ref.~\cite{Dermisek:2012as, Dermisek:2012ke}. With zero Yukawa couplings of vectorlike fermions the evolution of gauge couplings  only depends on the geometric means of  masses of vectorlike fermions with the same quantum numbers. Fixing two masses,  $M_{L_1}$ and $ M_{E_1}$, to 150 GeV is compensated by making masses of the other two vectrolike families correspondingly heavier. The addition of $\bar \lambda = 0.5$ does not change  the evolution of gauge couplings much since Yukawa couplings contribute only at the 2-loop level.
 
 The evolution of Higgs quartic coupling depends significantly on Yukawa couplings present in the model. In the standard model, the top Yukawa coupling already drives Higgs quartic coupling to negative values at a high scale. Additional sizable Yukawa couplings accelerate this behavior and thus the stability of the EW minimum sets a limit on the size of extra Yukawa couplings.
 
In the case of the SM extended by 3 vectorlike families the RG evolution of Higgs quartic coupling is significantly different. The Higgs quartic coupling can remain positive all the way to the GUT scale even with additional Yukawa couplings. 
  The difference  comes from larger values of all gauge couplings compared to the SM above the scale of vectorlike fermions. Larger gauge couplings slow down the running of Higgs quartic coupling, see Fig.~\ref{fig:RG}, and eventually turn the beta function of Higgs quartic coupling positive. This effect is further amplified by the fact that  the top Yukawa is driven fast to much smaller values compared to the SM (again due to larger gauge couplings)  and its contribution to the running of Higgs quartic coupling becomes small. 
 
 This is the minimal scenario (in this framework) with gauge coupling unification, sufficiently stable proton, and the Higgs quartic coupling remaining positive all the way to the GUT scale that simultaneously explains the deviation in the muon magnetic moment and generates the correct muon mass. 
 Adding another lepton Yukawa coupling of the same size, for example $\lambda$ to also modify $h\to \gamma \gamma$, would make the Higgs quartic coupling briefly go negative. The EW minimum would  still be sufficiently long lived even with somewhat larger values of lepton Yukawa couplings. Alternatively, a stable EW minimum  can be obtained (Higgs quartic coupling is positive all the way to the GUT scale) by lowering both extra lepton Yukawas to $\sim 0.4$.

 \section{Conclusions }
\label{sec:conclusions}

We showed that the deviation of the measured value of the muon anomalous magnetic moment from the standard model prediction can be completely explained by mixing of the muon with extra vectorlike leptons, L and E,  near the electroweak scale. This mixing simultaneously contributes to the muon mass (we label this contribution by $ m_\mu^{LE}$), and the correlation between contributions to the muon mass and muon g-2 is controlled by the mass of the neutrino originating from the  doublet L,  that is given by the vectorlike mass parameter $M_L$.

We have found two generic solutions: the asymptotic one,  $M_L \gg M_Z$, in which case the Higgs loop dominates and the measured value of the muon g-2 is obtained for $ m_\mu^{LE}/m_\mu \simeq -1$; and the second one with a light extra neutrino, $M_L \simeq M_W$, in which case the $W$ loop dominates and the measured value of the muon g-2 is obtain for $ m_\mu^{LE}/m_\mu \simeq +1$. In the first case, about twice as large contribution from the direct Yukawa coupling of the muon is required to generate the correct muon mass, while in the second case, the muon mass can fully originate from the mixing with heavy leptons. 

The sizes of possible contributions to the muon g-2, muon mass  and other observables depend on the upper limit on Yukawa couplings that we allow in the model.
We considered two cases, the upper limit being 0.5 and 1. The 0.5 upper limit is motivated by a simple UV embedding of this scenario  with three complete vectorlike families featuring gauge coupling unification, sufficiently stable proton, and the Higgs quartic coupling remaining positive all the way to the GUT scale. 

With the upper limit on Yukawa couplings being 0.5, the muon g-2 can be explained within one standard deviation either with  $M_L \lesssim 130$ GeV (the mass of the lightest extra charged lepton is $m_{e_4} \lesssim 150$ GeV), or with  $M_L \gtrsim 1$ TeV. The small $M_L$ case predicts the $h \to \mu^+ \mu^-$ in the range 5 -- 9 times the standard model prediction.
Depending on additional Yukawa coupling, the branching ratio for $h \to \gamma \gamma$
can be enhanced by   $\sim$15\% or lowered by $\sim$25\% from its SM prediction. The asymptotic case predicts only very small modifications of  $h \to \mu^+ \mu^-$ and $h \to \gamma \gamma$ compared to the SM.

Allowing Yukawa couplings of order 1, the range of parameters that can explain 
the muon g-2  within one standard deviation and the range of predictions for Higgs branching ratios significantly expand. The small $M_L$ case now requires  $M_L \lesssim 200$ GeV (the mass of the lightest extra charged lepton is $m_{e_4} \lesssim 250$ GeV). The predicted $h \to \mu^+ \mu^-$ ranges from 0.5 to 24 times the standard model prediction. A part of the parameter space is thus already excluded by the ATLAS search for this decay mode of the Higgs (the upper limit is 9.8 times the SM prediction).
Depending on additional Yukawa coupling, the $h \to \gamma \gamma$ rate can be modified by $\sim$50\%.

New vectorlike leptons generically predict a variety of flavor violating processes. The existing limits set strong constraints on other possible couplings in the model besides those needed for the explanation of the muon g-2 anomaly.  
In addition, extra charged leptons provide a variety of signatures at the LHC. They can be pair produced or can modify Higgs decays. Some searches are already excluding parts of the parameter space and others are getting close. Covering all possible decay modes of extra leptons should allow us to fully explore  the small $M_L$ case at the LHC with already available data.

\acknowledgements


R.D. thanks H.D. Kim for useful discussions. A.R. thanks K. Kannike for comparing the results of calculations.
R.D. also thanks Seoul National University and the KITP UCSB for kind hospitality during final stages of this project. This work was supported in part by the Department of Energy under grant number DE-FG02-91ER40661.



\end{document}